\documentclass{aa}  

\usepackage{graphicx}
\usepackage{txfonts}
\usepackage{color}
\begin{document} 

\title{The luminosity evolution of nova shells}

\subtitle{I. A new analysis of old data}

\author{C. Tappert\inst{1}
\and
N. Vogt\inst{1}
\and
A. Ederoclite\inst{2}
\and
L. Schmidtobreick\inst{3}
\and
M. Vu\v{c}kovi\'c\inst{1}
\and
L. L. Becegato\inst{2}
}

\institute{Instituto de F\'isica y Astronom\'ia,
Universidad de Valpara\'iso, Valpara\'iso, Chile\\
\email{claus.tappert@uv.cl}
\and
Instituto de Astronomia, Geof\'isica e Ci\^encias Atmosf\'ericas,
Universidade de S\~ao Paulo, S\~ao Paulo, Brazil
\and
European Southern Observatory
Santiago, Chile
}

\date{Received XXX; accepted XXX}

\abstract{
Over the last decade, nova shells have been discovered around a small number 
of cataclysmic variables that had not been known to be post-novae, while 
other searches around much larger samples have been mostly unsuccessful. 
This raises the question about how long such shells are detectable after the eruption and whether this time limit 
depends on the characteristics of the nova. So far, there has been
only one comprehensive study of the luminosity evolution of nova
shells, undertaken almost two decades ago. 
Here, we present
a re-analysis of the H$\alpha$ and [O{\sc iii}] flux data from that
study, determining the luminosities while also taking into account
newly available distances and extinction values, and including additional
luminosity data of `ancient' nova shells.
We compare the long-term 
behaviour with respect to nova speed class and light curve type. We find that,
in general, the luminosity as a function of time can be described as consisting
of three phases: an initial shallow logarithmic decline or constant behaviour, followed by a logarithmic main decline phase, with a possible return to
a shallow decline or constancy at very late stages.
The luminosity evolution in the first two phases is likely to be dominated by 
the expansion of the shell and the corresponding changes in volume and density,
while for the older nova shells, the interaction with the interstellar medium
comes into play.
The slope of the
main decline is very similar for almost all groups for a given emission line,
but it is significantly steeper for [O{\sc iii}], compared to H$\alpha$,
which we attribute to the more efficient cooling provided by the forbidden
lines. The recurrent novae are
among the notable exceptions, along with the plateau light curve type
novae and the nova V838 Her. 
We speculate that this is due to the presence of denser material, possibly
in the form of remnants from previous nova eruptions, or of planetary nebulae,
which might also explain some of the brighter ancient nova shells.
While there is no significant difference
in the formal quality of the fits to the decline when grouped according to
light curve type or to speed class, the former presents less
systematic scatter. It is also found to be advantageous in identifying
points that would otherwise distort the general behaviour.
As a by-product of our study, we revised the identification of all novae
included in our investigation with sources in the Gaia Data Release 2
catalogue.
}


\maketitle
%

\defcitealias{downesetal01-2}{D01}
\defcitealias{stropeetal10-1}{S10}

\section{Introduction \label{intro_sec}}

A nova eruption is an event in a cataclysmic variable (CV) that occurs
once the white dwarf (WD) has accreted a critical amount of material from its
late-type main-sequence star companion, triggering a thermonuclear runaway
on the former, resulting in a brightness increase by 8-15 mag 
and an explosive ejection of material into the interstellar medium
\citep{bode+evans12-1}. The typical mass of the ejected material, the nova 
shell, is estimated at $10^{-5}-10^{-4}$ M$_\odot$ \citep{yaronetal05-1},
and it 
is assumed that this represents roughly the amount of the previously accreted 
material. 
The explosion affects only the outer layers of the WD
and the CV itself is not destroyed in the process, in fact, it may even recommence 
mass-transfer within a couple of years afterwards \citep{retteretal98-2}. 
As a  consequence, the nova eruption is
a recurrent process, with the majority of the
novae having estimated recurrence times of $\ge 10^4$ yr 
\citep{sharaetal12-3,schmidtobreicketal15-1,sharaetal18-2}.

CVs can thus be regarded as novae that are in-between eruptions. However, several
subclasses of CVs exist, distinguished by their physical parameters such as
mass, orbital period, mass-transfer rate, etc. 
\citep[e.g.][]{warner95-1,hellier01-1}. While some theoretical models have
tried to take into account the role of these parameters in the nova eruption
and its recurrence time \citep[e.g.][and references therein]%
{townsley+bildsten05-1,sharaetal18-2}, the corresponding observational data
are still very scarce. The same is true for the consequence of the nova 
eruption for the evolution of CVs, where theoretical predictions
\citep[e.g. by][]{sharaetal86-1,schreiberetal00-2,schreiberetal16-1} cannot
be tested due to the lack of observational data. In this context, it appears that it is 
important to identify and study `ancient' novae in order to investigate
the properties of CVs long after they have experienced a nova eruption, so that  the
short-term effects of the eruption can be distinguished from potential long-term ones that may be present.

The smoking gun for establishing a CV as a former nova is the presence of 
a nova shell. 
This has now been successfully achieved for ten objects,  a few
well-known CVs among them (see Section \ref{an_sec}). Non-detections span slightly more than 110 CVs 
\citep{sahmanetal15-1,schmidtobreicketal15-1,pagnotta+zurek16-1}.
Even more puzzling is the absence of shells in post-novae: narrow-band
studies indicate that only $\sim$47\%\  of all 
novae actually present shells \citep{cohen85-1,downes+duerbeck00-1}.
All novae necessarily eject material. We consider, then, why  
 in about half of all novae, this material  emits light for a 
sufficient amount of time to be detected as an extended shell, while 
in the other half, any emission from a possibly forming shell fades 
away too quickly for detection.

To our knowledge, the only comprehensive study on the long-term behaviour
of the luminosities of nova shells was conducted by 
\citet[][hereafter \citetalias{downesetal01-2}]{downesetal01-2}, who 
investigated the evolution of the hydrogen and the [O{\sc iii}] $\lambda$500.7
nm luminosities of the 
shells by comparing novae of different ages as a function of the speed class, 
which are defined by the rate of photometric decline of the nova, measured as 
the time range in which the nova has declined from maximum by 2 ($t_2$) or 3 
($t_3$) magnitudes \citep{mclaughlin45-2,payne-gaposchkin64-1}. While that 
study represents an important and valuable step in the right direction, it has 
the following significant shortcomings \citep[see also][]{tappertetal17-1}:
1) Some speed classes are severely undersampled. 2) About half of the novae are
registered with a single data point only. Hence, grouping them into classes is 
necessary for deriving any trends. 3) Grouping according to speed class 
presupposes that this is the dominant parameter and prevents a proper parameter 
study.\footnote{In this context, 
it is worth mentioning that the speed class in some cases is ambiguous 
because it strongly depends on whether the nova had actually been observed at
maximum light,
on the completeness of the early light curve, or even on the interpretation
of the decline light curve by different authors.
One of the most
striking examples is the case of DK Lac, which \citetalias{downesetal01-2} list
with $t_3 = 24$ d, while \citet{stropeetal10-1} find $t_3 = 202$ d and 
\citet{selvelli+gilmozzi19-1} measure $t_3 = 60(15)$ d.}
4) The plots do not distinguish between individual systems, which makes it
impossible to directly identify objects that dominate the plot or those that
systematically deviate from the general trend
(since the data themselves are included in the article, this is a 
comparatively minor point).
5) The data are lacking any error estimation.

Our present study aims at improving some of those points. We divide it into
two parts. In the first, we revise the available data from 
\citetalias{downesetal01-2}, using distances from Gaia Data Release 2 
\citep{gaia16-1,gaia18-1} and interstellar reddening values from 
\citet{oezdoenmezetal16-2}, both of which include error estimation,
to calculate the corresponding luminosities. The undersampling of the data
for individual novae still makes it necessary to sort them into groups.
However, in addition to grouping according to speed class, we use the 
light curve types defined by 
\citet[][hereafter \citetalias{stropeetal10-1}]{stropeetal10-1} as a second
sorting criterion. In a second, forthcoming, part, we will present new flux 
values of nova shells that will add a time interval $\ge$20 yr to the 
\citetalias{downesetal01-2} data, and for some novae, yield a second data
point, so that the luminosity evolution can be studied for a larger sample
of individual objects.

\section{The data \label{data_sec}}

\subsection{The sample \label{sample_sec}}

\begin{figure}
\includegraphics[width=\hsize]{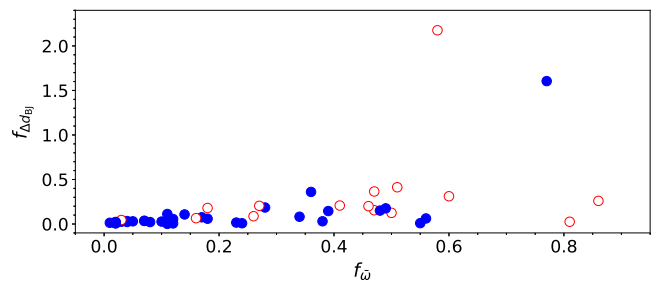}
\caption{Fractional difference between the inverse of the Gaia DR2
parallax and the distance from \citet{bailer-jonesetal18-4}, 
$f_{\Delta d_\mathrm{BJ}}$, as a function of the fractional error of the
parallax $f_{\bar{\omega}}$. Filled circles indicate novae with known light
curve type \citep{stropeetal10-1}, open circles indicate those without.}
\label{distancesigma_fig}
\end{figure}

\begin{figure}
\includegraphics[width=\hsize]{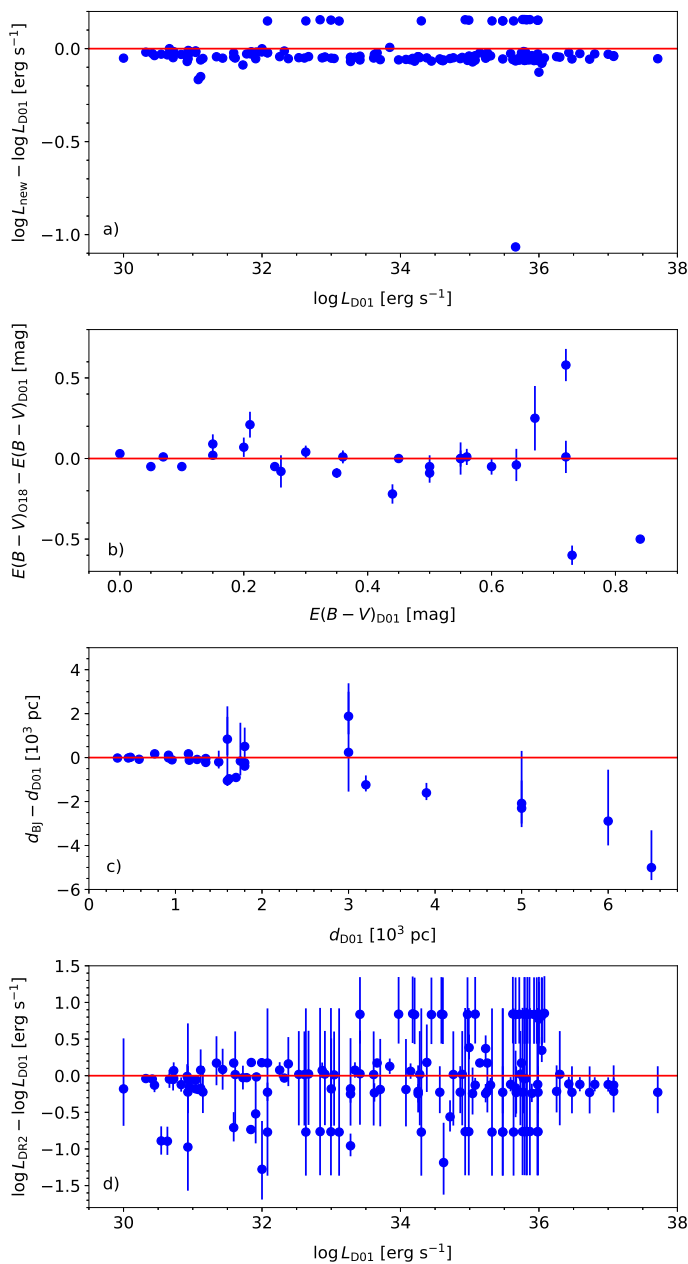}
\caption{Comparison with the parameters used in \citetalias{downesetal01-2}.
From top to bottom, the individual plots show a) the difference between the 
luminosities $L_\mathrm{new}$ calculated with Eq.\ref{lum_eq} using reddening 
and distances from \citetalias{downesetal01-2} and the luminosities 
$L_\mathrm{D01}$, b) the difference between the reddening values 
$E(B-V)_\mathrm{O18}$ from \citet{oezdoenmezetal18-1} and $E(B-V)_\mathrm{D01}$,
c) the difference between the distances $d_\mathrm{BJ}$ from 
\citet{bailer-jonesetal18-4} and $d_\mathrm{D01}$ , and d) the differences
between the luminosities $L_\mathrm{DR2}$ calculated using 
$E(B-V)_\mathrm{O18}$ and $d_\mathrm{BJ}$ with $L_\mathrm{D01}$.
All differences are plotted versus the respective values from 
\citetalias{downesetal01-2}. }
\label{d01parcomp_fig}
\end{figure}

The \citetalias{downesetal01-2} catalogue lists the shell luminosities for
the forbidden [O{\sc iii}] emission as well as for the hydrogen transitions
H$\alpha$ and H$\beta$. In this work, for the sake of brevity and focus, we 
do not consider the H$\beta$ line because it will track basically the same,
optically thicker, part of the shell as H$\alpha$, in contrast to [O{\sc iii}],
which will correspond to optically thinner material. We chose H$\alpha$ over 
H$\beta$ because it is the stronger line, especially at later stages, and the 
corresponding flux measurements would be less affected by noise.
The disadvantage of this choice is that the flux measured from a narrow-band 
filter centred on H$\alpha$ would potentially also include emission from the 
nearby [N{\sc ii}] lines at $\lambda$654.8 and $\lambda$658.3 nm. However, 
for the vast majority of the nova shells, the combined flux of the [N{\sc ii}]
lines would still amount to less than the H$\alpha$ emission, so that the total 
flux would differ from the pure H$\alpha$ flux by less than a factor of two, 
which, for our purposes, is not relevant. Still, we should be aware that when in
the following we talk about H$\alpha$ fluxes and luminosities, this actually
means H$\alpha$ + [N{\sc ii}].

The luminosity $L_\lambda$ of an emission at wavelength $\lambda$ can be
calculated from the flux $F_\lambda$ as

\begin{equation}
L_\lambda = 4 \pi d^2 F_\lambda~10^{\frac{c_\lambda R E(B-V)}{2.5}},
\label{lum_eq}
\end{equation}

where $d$ is the distance to the observer, $E(B-V)$ the reddening parameter, 
$R = 3.1$ the extinction law parameter \citep{cardellietal89-1}, and $c_\lambda = A_\lambda / A_V$
is the conversion factor between the wavelength specific absolute extinction $A_\lambda$ and
the one in the visual range, $A_V$. 
We used the York Extinction 
Solver\footnote{\url{http://www.cadc-ccda.hia-iha.nrc-cnrc.gc.ca/community/YorkExtinctionSolver/}} 
\citep{mccall04-1}, employing the reddening law from
\citet{fitzpatrick99-1}, to obtain the values for the two lines used in this
work to $c_\mathrm{H\alpha} = 0.75$ and $c_\mathrm{[O\,III]} = 1.111$.

Apart from the luminosities, the \citetalias{downesetal01-2} 
catalogue also provides information on the $d$, $E(B-V),$ and $F_\lambda$
values used for their calculation, however, as mentioned above, without including their 
associated uncertainties. 
With respect to the fluxes, only a few of the original source articles
include error estimations \citep[e.g.][]{ringwaldetal96-3}.
Additionally, a considerable part of the 
data is exclusive to the catalogue or is cited as `private communication'.
Thus, we made the decision  to take all flux data from 
\citetalias{downesetal01-2} to avoid giving the impression that certain data 
points are more precise than others. 

For the other two parameters, however, new measurements exist that also
provide an estimation of the associated uncertainties. For the reddening
data, we turned to the catalogue of \citet{oezdoenmezetal18-1}, where available,
while for the distances, we consulted the catalogue of \citet{bailer-jonesetal18-4}, which is based on the Gaia Data Release 2 (hereafter DR2) parallaxes
\citep{gaia18-1}. For a few objects,  there were also 3D reddening data  available 
from the DR2 data via the Stilism
website\footnote{\url{https://stilism.obspm.fr/}} \citep{lallementetal19-1}, 
which were found to be in agreement with those of \citet{oezdoenmezetal18-1}
within the errors.

The DR2 distances require some closer inspection. First of all, the novae
used by \citetalias{downesetal01-2} have to be cross-matched with the DR2
catalogue. This is less trivial than would otherwise be expected as most novae are
located in the crowded regions of the Galactic disc, and even queries 
with radii as small as 2 arcsec can give ambiguous results. We have 
therefore compared each Gaia source with the available finding charts and,
in a few cases, we used additional data to ensure the validity of a nova
identification. Details on this process and the results for each nova
included in \citetalias{downesetal01-2} are given in Appendix 
\ref{novapar_sec}.

Secondly, also the distances themselves have to be evaluated beyond taking into
account their formal uncertainties. The root of the problem here is the
non-Gaussian distribution of the latter that results when calculating the
distance as the inverse of the measured parallax $\bar{\omega}$, which
becomes more pronounced and skewed the larger the uncertainty of the parallax 
measurement is in comparison to the value of the parallax. This not only 
affects the estimation of the uncertainty associated with the distance, but 
the distribution will also yield a most probable value for the distance that
does no longer correspond to the true value \citep{lurietal18-1}. This 
problem can be addressed by interpreting the results in the context of
a more realistic model of the distance distribution, for example, by using a
prior according to Bayes' theorem \citep{bailer-jones15-2}, with the
recommended approach being to assume an exponentially decreasing space
density \citep[EDSD;][]{lurietal18-1}. This was done by 
\citet{bailer-jonesetal18-4}, who used a length scale map based on a
chemo-dynamical model of the Galaxy \citep{rybizkietal18-1} to estimate
the distances of the Gaia DR2 sources, and by \citet{schaefer18-2}, who
used a less complex approach of a length scale model as a function of
the galactic latitude to derive distances for a sample of 64 novae.

Still, we are not out of the woods yet because the resulting distances
continue to be affected by the uncertainties of the parallax measurements
$\sigma_{\bar{\omega}}$ in a non-uniform way. \citet{bailer-jones15-2}
has shown that for fractional errors $f_{\bar{\omega}} \equiv 
\sigma_{\bar{\omega}} / \bar{\omega} > 0.373$ the assumed prior starts to 
dominate the distance
estimation, that is,~the resulting value is more determined by the assumed model
than by the actual parallax measurement. Thus, to be on the safe side, we
should limit our sample to novae with fractional errors below this value. 
However, we have to consider that this could diminish the size of our sample
to a number where it loses statistical significance. In addition, the
luminosity range that is covered by the data spreads over several orders of
magnitude \citepalias{downesetal01-2}, so that even a very uncertain distance
might still prove useful for our purposes. Thus, we decided to define our
sample based on the Gaia DR2 information applying three criteria.

Firstly, we exclude all objects with negative parallaxes because while they
can be formally used to calculate distances, a comparison with data for
clusters shows that such distances present a systematic offset
\citep{bailer-jonesetal18-4}. Secondly, for similar reasons, we exclude
distances with fractional errors, $f_{\bar{\omega}} \ge 1,$ because these
allow for negative or zero parallaxes within the uncertainties. Thirdly,
in order to estimate the influence of the prior on the distance value, we
compare the distance, $d_{\bar{\omega}}$ , calculated as the inverse of the
parallax to the distances computed with a Bayesian prior, $d_\mathrm{BJ}$ and
$d_\mathrm{S}$, from \citet{bailer-jonesetal18-4} and \citet{schaefer18-2}, 
respectively, by defining absolute fractional differences, namely,
\begin{equation}
f_{\Delta d_\mathrm{BJ}} \equiv |d_\mathrm{BJ}-d_{\bar{\omega}}| /
d_{\bar{\omega}},
\end{equation}
and $f_{\Delta d_\mathrm{S}}$ , correspondingly. Figure~\ref{distancesigma_fig} 
shows $f_{\Delta d_\mathrm{BJ}}$ as a function of $f_{\bar{\omega}}$. For
clarity, we have omitted the values for V1419 Aql, which has $f_{\bar{\omega}}
= 0.33$ and $f_{\Delta d_\mathrm{BJ}} = 16.1$, and actually represents a good 
example for the fractional error alone not being a sufficient quality criterion.
Based on that plot, we set our limit to $f_{\Delta d_\mathrm{BJ}} < 0.5$,
since below that value, the distribution is comparatively uniform for
$f_{\bar{\omega}} > 0.2$. This means that the use of a prior does not result
in a distance difference larger than a factor of two compared to not using a prior
and also means that the corresponding luminosity is affected by a factor of $\le$4, which 
should still represent a reasonable uncertainty considering the luminosity 
spread. While not shown here explicitly, we performed the same comparison 
with the \citet{schaefer18-2} distances and obtained similar results. 
Specifically, we found that all objects with $f_{\Delta d_\mathrm{BJ}} < 0.5$ 
also have $f_{\Delta d_\mathrm{S}} < 0.5$ and vice versa, 
which proves the consistency of this limit.

For several objects, we applied additional criteria. The recurrent novae
RS Oph and V3890 Sgr formally are within our limits. However, both objects
contain giant secondary stars and have comparatively long periods.
\citet{schaefer18-2} argues that the resulting orbital wobble is of the same
size as the measured parallax, thus rendering the latter inapplicable. 
Because there is a decent amount of luminosity data available for these two 
objects,
and because the behaviour of the 
flux data within themselves for a particular system is independent of the
distance, we  included these objects for comparison, but we ought to keep in mind that
the zero point on the luminosity axis for those two objects is undetermined.
As for the other recurrent nova, T Pyx, there is only one very late data point,
whose placement on the time axis is uncertain since
\citet{schaeferetal10-1} argue that the observed shell is a remnant of an
eruption in the year 1866, instead of the one in 1967. In our analysis, we
therefore treat these objects with the appropriate caution and we do not
attempt to fit the data.

Finally, the nova V838 Her with $f_{\bar{\omega}} = 1.53$ does not qualify. 
However, as we will see below, there is no reasonably possible distance value 
that could reconcile its luminosity 
decline with any of the other objects, and so it will be discussed 
individually, 
making the assigned distance less important. 
In contrast to the cases of RS Oph and V3890 Sgr, here the distribution of the 
points is such that it allows
for a meaningful linear fit.
These steps resulted in a sample of 42 objects with associated DR2 distances from the
\citet{bailer-jonesetal18-4} catalogue and then further reduced to 29 objects 
which have a light curve type assigned (see Section \ref{evol_sec}).

In order to evaluate the consequences of employing these new sets of
parameters, we compared them to those listed in 
\citetalias{downesetal01-2}.
Firstly, however, we calculated the H$\alpha$ luminosities using the 
\citetalias{downesetal01-2} parameters to check for potential inconsistencies 
in the catalogue itself. In the top plot of Fig.~\ref{d01parcomp_fig} we
present the difference between  such newly calculated luminosities, 
$L_\mathrm{new}$ ,using Eq.\ref{lum_eq} and the values $L_\mathrm{D01}$ listed
in the catalogue as a function of the latter. We see that the bulk of the
points lies slightly below the zero line, which can be easily explained by
the use of a different reddening law and corresponding conversion factor, 
$c_\mathrm{H\alpha}$, and the scatter within this bulk is likely due to 
rounding. For deviating points farther off this bulk, we can only speculate 
that this is because the parameters used for the calculation do not correspond
to those listed in the paper, perhaps due to mistakes in the transcription. 
This would either concern reddening or distances, affecting all data points of 
a particular object (e.g. the row of points at $\sim$0.15 on the y-axis 
corresponds to all the V2214 Oph data)
or individual data (e.g. the most 
deviating point corresponds to one out of 25 data points for V992 Sco
and is clearly a typographical mistake in the exponent of the stated 
luminosity).

In the upper middle plot, we show the comparison with the reddening values 
$E(B-V)_\mathrm{O18}$ from \citet{oezdoenmezetal18-1}. 
We find that for $E(B-V)_\mathrm{D01} \le 0.6$ mag, they match 
well, while for stronger reddening, there is a considerably larger scatter.
The strongest deviating points with absolute differences $>$0.5 mag are
those of V888 Cen, BY Cir, and V992 Sco. For the first two of those, 
\citetalias{downesetal01-2} have used the maximum magnitude vs rate of decline 
relationship (MMRD) for classical novae to derive distance and reddening. 
However, this method, at  least in its past form, is 
considered unreliable today \citep[e.g.][]{schaefer18-2,selvelli+gilmozzi19-1}. 
For V992 Sco, \citetalias{downesetal01-2} also quote their own work, but they do 
not specify how the reddening was obtained. In all three cases, we thus 
consider the \citet{oezdoenmezetal18-1} reddening as more trustworthy.

Next, in the lower middle plot, we compare the distances. There is a
generally good agreement between the \citetalias{downesetal01-2} values
$d_\mathrm{D01}$ and those of DR2 for $d_\mathrm{D01} \le 1500$ pc, after 
which the scatter becomes larger, but still stays within the uncertainties of 
the DR2 data. For $d_\mathrm{D01} > 3000$ pc, the \citetalias{downesetal01-2}
values appear to be systematically larger than those of 
\citet{bailer-jonesetal18-4}. The largest deviation is observed for DO Aql,
which is an interesting case, since, as noted by \citetalias{downesetal01-2},
a proper application of the MMRD would have resulted in an even larger distance
of 9.5 kpc, while a comparison with the faintest novae would place it at
3.6 kpc, the latter at least being comparable with the DR2
distance $d_\mathrm{BJ} = 1.5^{+1.7}_{-0.6}$ kpc within the uncertainties.
Finally, \citetalias{downesetal01-2} decided to take the average of the
two extremes. 

Concluding the comparison, the bottom plot of Fig.~\ref{d01parcomp_fig} 
presents the difference between the \citetalias{downesetal01-2} data and the 
new luminosities that were 
calculated using the reddening from \citet{oezdoenmezetal18-1} and the DR2 
distances from \citet{bailer-jonesetal18-4}. The differences between the
two sets of luminosities can be considerable, amounting up to around one order
of magnitude. From the other plots, and also from Eq.\ref{lum_eq}, it is clear
that these are mainly due to the differences in the distance estimates.

\subsection{Ancient novae\label{an_sec}}

As mentioned briefly in the introduction, there are now ten cases where the
presence of a nova shell has revealed a previous nova eruption in a CV where
the eruption itself was not observed. For the lack of a more precise and 
poignant term, we call those systems `ancient novae', although this should
not be taken strictly in the sense of an age indicator. Still, typically
these nova eruptions will have occurred in epochs that predate the known
nova eruptions and could, perhaps, be dated in some cases if a historical Far 
Eastern Guest star can be identified with a modern CV as classical nova 
candidate \citep{vogtetal19-1,hoffmann19-1}. Determining the luminosity of 
their shells would, thus, extend the
time axis beyond the \citetalias{downesetal01-2} data and provide important
information on the late stages of the luminosity evolution. In the following,
we inspect nine of these cases, focussing on the available data 
regarding age and shell luminosity, and using the DR2 distances from 
\cite{bailer-jonesetal18-4} and the Stilism reddening \citep{lallementetal19-1} 
in the process (see Tables \ref{allD01_tab} and \ref{novaprop_tab}). We note that not all systems can be used for our purposes, but we still include 
them for completeness. The tenth ancient nova is a member of the globular
cluster M22 \citep{goettgensetal19-1}. Because of the possibility that the
formation and evolution of CVs in globular clusters is particular to the
conditions in the cluster \citep{knigge12-1,bellonietal16-2,bellonietal17-1,bellonietal17-2,bellonietal19-1} and that they
are not representative, thus, for the general population of CVs, we do not 
 consider such objects further.

\subsubsection{V1315 Aql}

This is a nova-like variable of the SW Sex sub-type \citep{hellier96-4},
which is suspected to harbour the CVs with the highest mass-transfer rates
\citep{rodriguez-giletal07-1,townsley+gaensicke09-1,schmidtobreick15-1}. 
The shell was detected by \citet{sahmanetal15-1} in an H$\alpha$ survey.
A preliminary age estimate of 120 yr was later corrected to a possible
age range of 500 to 1200 yr, strongly depending on the original velocity of
the ejecta \citep{sahmanetal18-1}. 
Here, we will assume a mean value of 850 $\pm$ 350 yr. The authors
calculate a total flux for H$\alpha$ to $F =2.5(6) \times 10^{-13}$ erg 
cm$^{-2}$ s$^{-1}$. From their Table 4, we can estimate that the combined flux
of the neighbouring [N{\sc ii}] lines amounts to roughly 2/3 of that value, so 
that a typical narrow band filter centred on H$\alpha$ would yield an overall 
flux of about $4.2(7) \times 10^{-13}$ erg cm$^{-2}$ s$^{-1}$, implying 
$L = 1.3^{+0.5}_{-0.4} \times 10^{31}$ erg s$^{-1}$.

\subsubsection{V341 Ara}

V341 Ara is one of the brightest CVs on the sky, but it was only 
 recognised as such relatively recently \citep{frew08-1,bond+miszalski18-1}. It is 
associated to an H$\alpha$ emission
nebula and the authors referenced above list a nova eruption as one of the potential
scenarios, with the proper motion data implying an age of the nova of about
800 yr. Adding up the fluxes from [N{\sc ii}] and H$\alpha$ given by
\citet{frew08-1} yields a combined flux $F = 6.4 \times 10^{-14}$
erg cm$^{-2}$ s$^{-1}$. Using the DR2 and Stilism data on distance and
reddening distribution, respectively, yields $L = 2.0^{+0.2}_{-0.1} \times
10^{29}$ erg s$^{-1}$. For [O{\sc iii}] the corresponding values are
$F = 5.5 \times 10^{-14}$ erg cm$^{-2}$ s$^{-1}$ and
$L = 1.8^{+0.2}_{-0.1} \times 10^{29}$ erg s$^{-1}$.

\subsubsection{Z Cam}

This was the first CV unmasked as an ancient nova \citep{sharaetal07-1} and it
is still the oldest, with an estimated age of 1300 to $\ge$5000 yr
\citep{sharaetal12-3}. It has been suggested that it can be identified with a 
guest star from the year 77 BCE as recorded by Chinese astrologers 
\citep{johansson07-1}, but this has been recently disputed by
\citet{hoffmann19-1}. Unfortunately, the Z Cam shell so far has only been 
analysed in
terms of extension and expansion, but not in flux, so that it cannot be
used in the context of the present paper.

\subsubsection{BZ Cam}

The most recently found coincidence of a CV with a possible nova shell refers 
to this nova-like variable of VY Scl sub-type with an orbital period of 3.7 h
\citep{honeycuttetal13-1}. \citet{hoffmann+vogt20-sub} report that BZ Cam
could be the counterpart of an  ancient guest star observed by Chinese 
astronomers in the year 369 CE, and that it seems to be associated to the 
faint nebula EGB 4, a planetary nebula candidate with a multiple 
structure \citep{ellisetal84-1}. BZ Cam is situated at the edge of a red 
nebula (apparently caused by H$\alpha$ emission) and surrounded by a smaller 
bluish [O{\sc iii}] nebula, implying a nova shell. Based on the DR2 distance 
and proper motion values of BZ Cam,  \citet{hoffmann+vogt20-sub} estimate that 
the shell structure could have witnessed a total of three past ejection 
episodes, about two, five and eight millennia ago. It could be the first 
candidate for a recurrent nova with extremely long repetition cadence.

\citet{greineretal01-9} estimated the total [O{\sc iii}] flux of the nebula
to $F = 4.8 \times 10^{-13}$ erg cm$^{-2}$ s$^{-1}$, while 
\citet{krautteretal87-2} gave $F = 4.3 \times 10^{-13}$ erg cm$^{-2}$ s$^{-1}$
and $F = 1.3 \times 10^{-13}$ erg cm$^{-2}$ s$^{-1}$ for the eastern and
western part of the nebula, respectively, which, in total, thus yields a slightly
higher value, but stays within the same order of magnitude. The spatial
distribution of the H$\alpha$ and [O{\sc iii}] fluxes is highly inhomogeneous.
From the data given in \citet{hollisetal92-3}, \citet{krautteretal87-2} and
\citet{greineretal01-9}, we estimate a rough average of 
$F(\mathrm{[OIII]}) / F(\mathrm{H}\alpha+\mathrm{[NII]}) \sim 1.7$ for the
flux ratio of the lines relevant for our study. With the
 flux value for [O{\sc iii}] from \citet{greineretal01-9}  given above and their stated uncertainty of 20\%, we thus estimate a corresponding flux
for H$\alpha$ to $F = 2.8(6) \times 10^{-13}$ erg cm$^{-2}$ s$^{-1}$,
yielding luminosities of $L = 5.1^{+1.9}_{-1.5} \times 10^{30}$ erg 
s$^{-1}$ for H$\alpha$ and $L = 9.1^{+3.9}_{-3.0} \times 10^{30}$ erg s$^{-1}$
for [O{\sc iii}].

\subsubsection{AT Cnc}

The shell around this Z Cam sub-type dwarf nova was discovered by
\citet{sharaetal12-4} and later analysed in more detail by 
\citet{sharaetal17-2}. The
authors estimate the age of the nova as 330$^{+135}_{-90}$ yr. The emission
of the shell is dominated by forbidden transitions, with very little 
contribution from hydrogen, if any. The flux measured with a narrow band
filter on H$\alpha$ will thus mostly correspond to the $\lambda$654.8 and 
$\lambda$658.3 nm [N{\sc ii}] emission. Adding up the flux of all the
emission blobs measured by \citet{sharaetal17-2} yields 
$F = 3.3(3) \times 10^{-14}$ erg cm$^{-2}$ s$^{-1}$. The corresponding 
luminosity then calculates to $L = 8.5 \pm 1.3 \times 10^{29}$ 
erg s$^{-1}$.

\subsubsection{V1363 Cyg}

\citet{sahmanetal15-1} present an H$\alpha$ image that shows possible traces
of a nova shell centred on this dwarf nova, with its proximity to a gas cloud
impeding an unambiguous identification. To our knowledge, no further
analysis of this potential shell has been undertaken so far. While there is
thus no further use for this system within the scope of our study, we allow
ourselves a brief tangent, taking advantage of the availability of a
good distance measurement. The projected radius of the potential shell is 
estimated to 60 arcsec \citep{sahmanetal15-1}. With $d_\mathrm{BJ} =
1.7 \pm 0.1$ kpc, this translates to an absolute extension 
$r = 0.49 \pm 0.03$ pc. In the attempt to estimate the age of a nova shell,
it is often considered that an original expansion velocity in the order of
$\sim$2000 km/s decreases exponentially with a half-life of $\sim$75 yr, which 
is based on a study of four nova shells by \citet{duerbeck87-4} and which
is thought to be caused by interaction with the surrounding interstellar
medium \citep{oort46-1}. In the case of V1363 Cyg, these values do not
agree with the observed extension of the shell, requiring either a
considerably higher ejection velocity $>$4000 km/s or a longer half-life
$>$100 yr. In a recent study of five shells, \citet{santamariaetal20-1} did 
not find any compelling evidence for deceleration, and if the feature in the
H$\alpha$ images of V1363 Cyg indeed corresponds to a nova shell, this might be
another example for above rule of thumb not being applicable in a general
way.

\subsubsection{CRTS J054558.3+022106}

This eclipsing dwarf nova was detected within the suspected planetary
nebula Te11, which, based on this discovery, was then reinterpreted as an
ancient nova shell \citep{miszalskietal16-1}. The authors identify the
nova eruption with a guest star recorded at the end of the year 483 CE,
which, according to \citet{hoffmann19-1}, is possible, but ambiguous.
\citet{miszalskietal16-1} report an H$\alpha$ flux
$F = 2.4 \times 10^{-13}$ erg cm$^{-2}$ s$^{-1}$, which is likely to represent
a close lower limit, because their aperture photometry did not encompass the
full nebula. From their Table 3, we estimate a flux for [O{\sc iii}] 
$F = 2.5 \times 10^{-13}$ erg cm$^{-2}$ s$^{-1}$. 
Unfortunately, the DR2 parallax is accompanied by a large
uncertainty of almost 1 mas. The corresponding distances are
$d_{\bar{\omega}} = 0.6^{+1.1}_{-0.2}$ pc and $d_\mathrm{BJ} =
1.0^{+1.2}_{-0.5}$ kpc, which yields $f_{\Delta d_\mathrm{BJ}} = 0.67,  $ which is, thus, above our quality limit for the distance and so, the luminosity
calculation (Section \ref{sample_sec}). \citet{miszalskietal16-1} derive
$d = 330(50)$ pc based on the flux and spectral type estimations of the
donor star in the dwarf nova system. However, they also give a reddening
$E(B-V) = 0.32$ mag\footnote{There is a typographical error at some point
in that paper, because the authors give a value of 0.38 mag in their section 
2.3, but refer to it as being derived in their section 3.2, where it is 
reported as 0.32 mag. Here we use the latter value.}, and according to
the Stilism data, this definitively implies $d > 550$ pc, with the values
presenting a comparatively narrow transition in the reddening distribution 
just below 500 pc, where $E(B-V)$ rises from $<$0.1 to $\ge$0.29 mag within 
50 pc. For a very rough estimate, we use the lower limit of 550 pc as implied 
by the reddening, and a maximum of 2.2 kpc as indicated by the $d_\mathrm{BJ}$
formal distribution to obtain $L \sim 1.1^{+1.7}_{-0.9} \times 10^{32}$
erg s$^{-1}$. The corresponding value for [O{\sc iii}] results to
$L \sim 1.6^{+2.4}_{-1.3} \times 10^{32}$ erg s$^{-1}$.

\subsubsection{IGR J17014$-$4306}

Another eclipsing dwarf nova and possible intermediate polar 
\citep{potter+buckley18-1}, this object has been associated by proper motion 
analysis with a nebula and a guest star sighting in the constellation of 
Scorpius recorded by Korean observers in 1437 CE \citep{sharaetal17-4}. The 
latter has been recently challenged by \citet{hoffmann19-1} who, based on a 
reanalysis of the historical texts, argues that the position of the guest star 
does not agree with the location of the nebula. However, since a comparison with 
the proper motion of the binary and the centre of the nebula yields a 
comparable time scale, here we  assume an age of $\sim$600 yr. This is 
sufficient for our purpose, considering that we compare on a 
logarithmic scale. \citet{sharaetal17-4} give a combined H$\alpha$ + 
[N{\sc ii}] flux $F = 2.8 \times 10^{-12}$ erg cm$^{-2}$ s$^{-1}$, which 
yields a luminosity of $L = 7.4^{+3.2}_{-2.2} \times 10^{32}$ erg s$^{-1}$.

\subsubsection{IPHASX J210204.7+471015}

This is a nova-like variable with a likely orbital
period of 4.3 h \citep{guerreroetal18-1}. \citet{santamariaetal19-1}
estimate the age of the shell to 147(20) yr. From Table 4 in 
\citet{guerreroetal18-1}, we derive a combined H$\alpha$ + [N{\sc ii}] flux
$F = 4.4 \times 10^{-14}$ erg cm$^{-2}$ s$^{-1}$, which, with the DR2 and
Stilism data, implies $L = 5.0^{+1.5}_{-1.2} \times 10^{30}$  erg s$^{-1}$.
Similarly, for [O{\sc iii}] $\lambda$500.7 nm, $F = 2.7 \times 10^{-15}$ erg 
cm$^{-2}$ s$^{-1}$ and $L = 4.1^{+1.8}_{-1.3} \times 10^{29}$  erg s$^{-1}$.

\section{Luminosity evolution \label{evol_sec}}

\begin{table}
\caption{Distribution of the novae in our sample among speed classes (rows)
and light curve types (columns). The final row and column gives the total
number of entries in a specific light curve type and speed class, respectively.}
\label{scltdis_tab}
\centering
\begin{tabular}{llllllll}
\hline
\hline
                 & D  & F & J & O & P & S & $n_\mathrm{SC}$ \\
\hline\noalign{\smallskip}
VF               & 1  & 0 & 0 & 2 & 1 & 2 & 6 \\
F                & 2  & 0 & 1 & 1 & 3 & 3 & 10 \\
MF               & 6  & 0 & 0 & 0 & 0 & 1 & 7  \\
SVS              & 0  & 2 & 4 & 0 & 0 & 0 & 6 \\
$n_\mathrm{LCT}$ & 9  & 2 & 5 & 3 & 4 & 6 & (29) \\
\hline
\end{tabular}
\end{table}

\begin{figure*}
\includegraphics[width=\hsize]{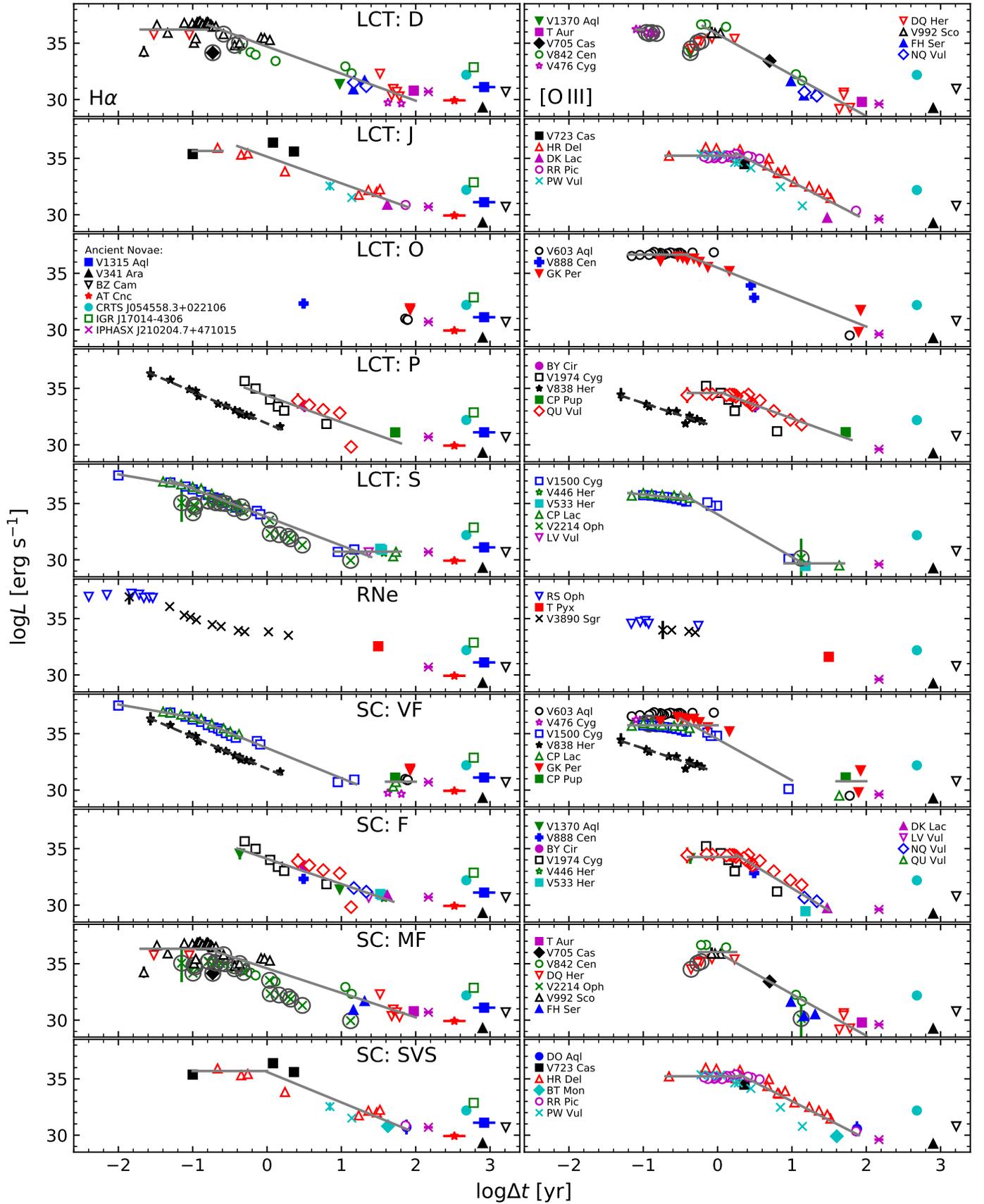}
\caption{H$\alpha$ (left panels) and [O{\sc iii}] (right panels) emission line
luminosities as a function of the time that has passed since maximum brightness.
Both parameters are shown in logarithmic scaling. The novae are grouped either
with respect to their light curve type (LCT) or speed class (SC). 
The ancient
novae have been incorporated in all plots, for comparison. 
The symbol 
identifications correspond to the respective plots of both emission lines. 
Symbols enclosed in larger grey circles indicate data points that were excluded
from the fit. Error bars are shown only for the first data point of a 
respective object. Solid light grey lines represent linear fits to the data of 
a specific group, the dashed darker grey line shows the fits to the V838 Her 
data.}
\label{lumevol_fig}
\end{figure*}

\begin{table}
\caption{Parameters $a$ and $b$ of the linear fits according to 
Eq.~\ref{fit_eq}. $\Delta(\log\Delta t)$ gives the time range of the 
corresponding segment, with $\Delta t$ in yr, $n$ is the number of 
data points used in the fit, and $\sigma$ is the standard deviation of 
the linear regression. For a few groups, two rows are given for the same
interval. The first corresponds to the linear fit, while the second
states the mean value for $b$, setting the slope $a$ to zero.
}
\label{fitpar_tab}
\centering
\setlength{\tabcolsep}{4pt}
\begin{tabular}{lr@{ .. }lllll}
\hline
\hline
Class & \multicolumn{2}{l}{$\Delta(\log\Delta t)$} & $n$ & $b$ & $a$ & $\sigma$ \\
\hline\noalign{\smallskip}
\multicolumn{7}{l}{H$\alpha$} \\
D        & $-$1.7 & $-$0.8 & 15 & 36.21(21) & 0           & 0.79 \\
         & $-$0.8 &    2.0 & 28 & 34.70(14) & $-$2.40(13) & 0.64 \\
J        & $-$1.0 & $-$0.6 &  2 & 35.65(27) & 0           & 0.27 \\
         & $-$0.4 &    1.9 & 13 & 35.14(35) & $-$2.36(31) & 0.81 \\
P        & $-$0.3 &    1.8 & 13 & 34.36(32) & $-$2.35(43) & 0.82 \\
S        & $-$2.0 & $-$1.0 &  8 & 35.32(17) & $-$1.13(13) & 0.11 \\
         & $-$2.0 & $-$1.0 &  8 & 36.76(13) & 0           & 0.35 \\
         & $-$1.3 &    1.4 & 24 & 33.79(06) & $-$2.52(06) & 0.22 \\
         &    0.9 &    1.8 &  8 & 30.73(07) & 0           & 0.20 \\
V838 Her & $-$1.6 &    0.2 & 13 & 31.89(07) & $-$2.83(09) & 0.14 \\
VF       & $-$2.0 & $-$1.0 &  8 & 35.32(17) & $-$1.13(13) & 0.11 \\
         & $-$2.0 & $-$1.0 &  8 & 36.76(13) & 0           & 0.35 \\
         & $-$1.0 &    1.2 & 19 & 33.72(06) & $-$2.67(08) & 0.21 \\    
         &    1.6 &    2.0 &  7 & 30.74(32) & 0           & 0.83 \\
F        & $-$0.4 &    1.7 & 22 & 34.11(23) & $-$2.24(24) & 0.69 \\
MF       & $-$1.7 & $-$0.6 & 21 & 36.33(15) & 0           & 0.70 \\
         & $-$0.7 &    2.0 & 18 & 34.59(24) & $-$2.17(19) & 0.71 \\
SVS      & $-$1.0 &    0.0 &  5 & 35.69(18) & 0           & 0.41 \\   
         &    0.0 &    1.9 & 12 & 35.61(49) & $-$2.70(38) & 0.78 \\
\hline\noalign{\smallskip}
\multicolumn{7}{l}{[O{\sc iii}]} \\
D        & $-$0.2  &    2.0 & 21 & 35.80(22) & $-$3.63(20) & 0.66 \\
J        & $-$0.7  &   0.25 & 23 & 35.23(06) & 0           & 0.28 \\
         &   0.25  &    1.9 & 23 & 36.30(31) & $-$3.39(32) & 0.72 \\
O        & $-$1.2  & $-$0.4 & 21 & 36.66(05) & 0           & 0.21 \\
         & $-$0.4  &    2.0 & 12 & 35.50(34) & $-$2.61(31) & 0.97 \\
P        & $-$0.4  &    0.1 &  6 & 34.60(12) & 0           & 0.29 \\
         &    0.1  &    1.8 & 26 & 34.73(16) & $-$2.39(27) & 0.50 \\
S        & $-$1.2  & $-$0.4 & 15 & 35.11(13) & $-$0.68(17) & 0.14 \\
         & $-$1.2  & $-$0.4 & 15 & 35.61(05) & 0           & 0.19 \\
         & $-$0.5  &    1.2 &  9 & 34.04(16) & $-$3.78(28) & 0.50 \\
         &    0.9  &    1.7 &  3 & 29.69(17) & 0           & 0.29 \\
V838 Her & $-$1.3  & $-$0.2 & 10 & 31.54(16) & $-$2.16(24) & 0.25 \\
VF       & $-$1.2  &    0.0 & 27 & 35.74(08) & 0           & 0.40 \\
         & $-$0.5  &    1.0 & 14 & 34.51(21) & $-$3.65(52) & 0.71 \\
         &    1.6  &    2.0 &  5 & 30.78(43) & 0           & 0.95 \\  
F        & $-$0.4  &   0.25 & 17 & 34.26(12) & 0           & 0.49 \\
         &   0.25  &    1.5 & 23 & 35.46(23) & $-$3.96(31) & 0.55 \\
MF       & $-$0.25 &   0.25 &  8 & 36.04(17) & 0           & 0.48 \\
         &    0.1  &    2.0 & 13 & 36.00(50) & $-$3.69(38) & 0.76 \\
SVS      & $-$0.7  &    0.3 & 26 & 35.23(06) & 0           & 0.31 \\
         &    0.3  &    1.9 & 21 & 36.37(34) & $-$3.35(31) & 0.68 \\
\hline
\end{tabular}
\end{table}

In their investigation of the luminosity evolution, \citetalias{downesetal01-2}
distinguished between different speed classes. In the present work, we also 
want to examine a possible correlation with the light curve type as defined
by \citetalias{stropeetal10-1}. To be able to conduct a proper comparison
for the two different groupings, we limit the sample to those objects in
\citetalias{downesetal01-2} that also have a light curve type assigned. 

The speed class system used by \citetalias{downesetal01-2} is that of
\citet{payne-gaposchkin64-1} who defined the following intervals for $t_2$:
up to 10 d (Very Fast, VF), 10 to 25 d (Fast, F), 26 to 80 d (Moderately Fast,
MF), 81 to 150 d (Slow, S) and from 151 d on (Very Slow, VS). 
\citetalias{downesetal01-2} do not strictly follow that scheme, but take, rather, $t_3$ 
as a second indicator whenever there is a significant break in the decline 
law. For example, V723 Cas, with $t_2$ = 19 d and $t_3$ = 180 d would be 
classified as Slow instead of Fast. However, this introduces some personal 
ambiguity (and furthermore emphasises the weakness of a scheme that tries to 
categorise nova light curves using one or two decline rate parameters). There 
are also some inconsistencies in the catalogue; for example, V446 Her with $t_2$ = 
7 d and $t_3$ = 12 d is classified as Fast, while GK Per with identical $t_2$, 
but $t_3$ = 13 d is labelled as Very Fast\footnote{Note, however, that
\citetalias{stropeetal10-1} find very similar parameters for GK Per, but 
$t_2$ = 20 d, $t_3$ = 42 d for V446 Her, which would agree with the class
assigned by \citetalias{downesetal01-2}, so perhaps this is just a
transcription error in the latter catalogue.}. To make a proper comparison possible,
we decided to use the speed classes assigned by \citetalias{downesetal01-2},
regardless of the actual decline rate parameters. Furthermore, we follow
their scheme of combining the two slow classes and hereafter, we refer to
the latter as SVS.

In contrast, \citetalias{stropeetal10-1} used purely phenomenological
criteria to characterise the decline light curves of novae and to define seven 
different types based on their distinctive features. In our sample, we found 
members of six of these types: D (Dust Dip), F (Flat Top), J (Jitter), 
O (Oscillations), P (Plateau), and S (Smooth). Only the least frequent
class, C (Cusp, which \citetalias{stropeetal10-1} find to include only
1\%\ of the novae), is not represented in our sample. 
However, we will 
also have to disregard the second least frequent one, F (2\%), because 
there are only two members (BT Mon and DO Aql), each of which counts with only 
a single data point per emission line.

In Table \ref{scltdis_tab}, we compare the distributions of the different
novae of our sample both for speed class and for light curve type to see
whether these two different groupings are indeed independent from each other.
We do find a rough correlation, with the O, P, and S light curve types
showing a concentration towards faster speed classes, while the D and J
types include slower novae.
We note that
although the small numbers per individual bin  suggests we should use some caution
when considering the significance of this result, it quite faithfully, in fact, reflects
the distribution in the \citetalias{stropeetal10-1} catalogue itself.
Thus, if the rate of decline of the broad-band photometric nova brightness 
(i.e.~the speed class) were the dominant parameter also for the shell
luminosity and its evolution, we would expect very similar behaviour 
for those light curve types that sample similar speed class regimes, 
that is,~ O, P, and S on the one side and D and J on the other.

The parameters that were used to calculate the luminosities as well as the grouping
criteria of all novae in our sample are given in the appendix in Table
\ref{novaprop_tab}. The resulting luminosities $L$ of the different groupings are shown 
in Fig.~\ref{lumevol_fig} as a function of the time $\Delta t$ that has passed since the
maximum brightness of the eruption. The errors propagate only from the uncertainties 
in the distance and reddening measurements and, thus, they are identical for each data point
for a given nova. Therefore, they do not affect the general tendency of specific systems
but they do indicate possible displacements of that tendency along the luminosity axis.
As a result, we  chose to include the error bars only in the first
data point of a nova in the plots.
Following the example of \citetalias{downesetal01-2}, we divide the distributions
into segments with linear behaviour on the logarithmic scale and fitted linear 
functions, 
\begin{equation}
\log L = a \log{\Delta t} + b
\label{fit_eq}
\end{equation}
to them, with $L$ in erg s$^{-1}$ and $\Delta t$ in yr. The limits of the 
$\log \Delta t$ range of a particular segment were determined by eye. 
In cases where the slope was consistent with
a zero value within the errors, it was set to zero, with the $y$-axis intersection 
$b$ corresponding to the average of the data points in that segment. In Table
\ref{fitpar_tab}, we give the details of the segments and the fit parameters. The
H$\alpha$ data for the O light curve type were found too sparse to yield a
meaningful fit. The results and their implications are discussed in Section 
\ref{disc_sec}. In the following, we remark briefly on a few noteworthy topics 
and objects.

\medskip\noindent
\textbf{Dust Dips:} The D light curve type is defined by presenting a significant
isolated minimum in the post-eruption light curve. This feature is caused by dust
absorption and has been found to also affect line emission \citep{shoreetal18-1}.
In our data, this is especially evident in the [O{\sc iii}] data of DQ Her (top right 
panel of Fig.~\ref{lumevol_fig}), where a comparison with the broad-band light curve
in \citetalias{stropeetal10-1} (their figures 2 and 7) shows that the first three of
our data points coincide with the final phases of the dust dip. Because of
the distorting effect on the general decline behaviour, time ranges corresponding 
to such broad-band photometric dust dips were excluded from the fits, both 
in the D type and in the speed class groupings. 
This concerns the novae V476 Cyg, DQ Her, and V992 Sco.

\medskip\noindent
\textbf{V1370 Aql:}
This nova also belongs to the D light curve type. However, the comparison with 
\citetalias{stropeetal10-1} (where it is mislabelled in the plots as
V1370 Cyg) shows that all data points lie clearly outside
the visual dust feature. Nevertheless,  the [O{\sc iii}] data point, in particular,
presented a significantly diminished luminosity compared to the rest of the D
group. Upon further investigation, we indeed found a discrepancy between
the flux values quoted by \citetalias{downesetal01-2} and those of the
original source corresponding to the time $\Delta t$ = 0.43 yr
\citep{snijdersetal87-1}. There, two flux values are given per emission
line that correspond to an identified broad and narrow component. For 
H$\alpha$, these are $F_b = 8.0 \pm 1.0 \times 10^{-12}$ erg cm$^{-2}$ s$^{-1}$
and $F_n = 2.0 \pm 0.5 \times 10^{-12}$ erg cm$^{-2}$ s$^{-1}$, respectively,
while the fluxes for [O{\sc iii}] $\lambda$500.7 nm are given as
$F_b = 2.0 \pm 0.5 \times 10^{-12}$ erg cm$^{-2}$ s$^{-1}$ and
$F_n = 0.2 \pm 0.1 \times 10^{-12}$ erg cm$^{-2}$ s$^{-1}$. In comparison,
\citetalias{downesetal01-2} state $F = 2.0\times 10^{-12}$ erg cm$^{-2}$ 
s$^{-1}$ for H$\alpha$ and $F = 2.9\times 10^{-13}$ erg cm$^{-2}$ 
s$^{-1}$ for [O{\sc iii}]. The only way that we can seek to reconcile these data
with the original source is the explanation that \citetalias{downesetal01-2} chose to
only consider the narrow components and, additionally, they made a mistake in the
transcription of the [O{\sc iii}] flux (e.g. 2.9 instead of 2.0). Yet even if 
that were the case, the choice for the narrow component is peculiar, 
considering that \citet{snijdersetal87-1} view the broad component as more 
likely to correspond to the ejected material. Because the grouping also puts 
together spectroscopic and photometric data, here we  decided to replace the
\citetalias{downesetal01-2} values with the combined fluxes from
\citet{snijdersetal87-1}.

Still, while in the Fast speed class groupings and also in the H$\alpha$
D type plot, the revised data now agree with the general distribution,
the [O{\sc iii}] luminosity remains too low by at least two orders of
magnitude when compared to the other D type members V476 Cyg and V842 Cen,
as it lies even slightly below the very first (dust) data point of DQ Her.
Because it clearly represents an outlier and would have a disproportionally
large effect on the fit, it was excluded.

\medskip\noindent
\textbf{\object{V838 Her}:} This nova is classified as a P light curve type 
and as a member of the very fast speed class. However, its luminosity slope 
deviates from the general behaviour in both groups. 
As shown in Section \ref{sample_sec}, the distance to V838 Her cannot
be reliably determined. However, from Fig.~\ref{lumevol_fig}, it is clear
that not only is the luminosity too low by about two orders of magnitude,
requiring a larger distance by a factor of 10 to correct that displacement,
but also that the decline starts significantly earlier than for the
other systems. Thus, there is no distance value that would allow to reconcile 
its behaviour with those of the other novae. 
The large amount of flux data for this object makes it possible to 
discuss it independently
(see Section \ref{hvso_subsec}). 

\medskip\noindent
\textbf{BT Mon:} 
\citetalias{downesetal01-2} do not assign any speed class to this nova. However, the
decline rate indicators $t_2$ and $t_3$ were determined by \citetalias{stropeetal10-1}
to 118 and 182 d, respectively, which clearly indicates a slow nova. We have included the
data points of BT Mon in the respective plots of this speed class, where it appears
to agree reasonably well with the general distribution.

\medskip\noindent
\textbf{V2214 Oph:} \citetalias{stropeetal10-1} place this nova among the S light curve types. However, looking at the broad-band light curve in their Figure 3, this object is
clearly the one\footnote{With the possible exception of the recurrent nova T CrB.} of the 32 
members of this class that fits the description of a smooth
decline the least, presenting instead a number of minima and maxima that deviate
significantly from the average decline law. As can be seen in the respective panel
in Fig.~\ref{lumevol_fig}, the H$\alpha$ data present a very similar behaviour. We
consider \object{V2214 Oph} as a very uncertain member of the S light curve class and thus we exclude
it from the corresponding fits.

\section{Discussion \label{disc_sec}}

\begin{figure}
\includegraphics[width=\hsize]{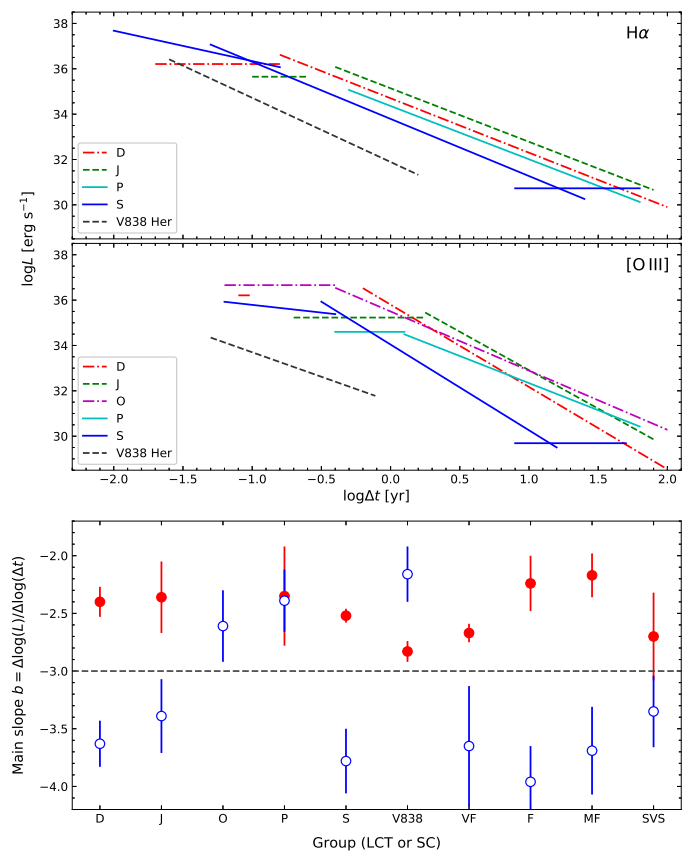}
\caption{Comparison of the fits as defined in Table \ref{fitpar_tab}. For
the sake of clarity, we omit the speed classes. The bottom plot compares
the main decline slopes of all groups (filled circles = H$\alpha$,
open circles = [O{\sc iii}]). The horizontal dashed line marks a slope of
$-$3.0.}
\label{fits_fig}
\end{figure}

\subsection{General description \label{gendesc_subsec}}

The interpretation of the data and the corresponding fits in Fig.~\ref{lumevol_fig}
call for a very cautious approach due to the fact that several regions
in these distributions are severely undersampled. While there appears to be
sufficient justification for separating the data into a number of sections and
to apply a linear fit to each of them (which, of course, represents only a 
first approximation to a necessarily smooth continuous function), the choice
of the corresponding time ranges is often ambiguous. Still, in general terms, 
the tendency is very similar for all groups. 
An initial, constant or very gentle
slope is followed by a main decline that for H$\alpha$ starts at
$\log \Delta t \le$0.0 (i.e.~$\le$1 yr after the eruption), and for [O{\sc iii}] 
at $\le$0.48 ($\le$3 yr). 
The S light curve type and the 
Very Fast Speed Class additionally present another late, approximately constant 
behaviour. 
However, we note that while it is only in these two groupings that this late stage
is obvious, in fact, the distributions in all groups are sufficiently ambiguous to allow, in principle, for the existence of such behaviour from 
$\log \Delta t \approx$1.5 (30 yr) onwards. 
\citetalias{downesetal01-2} also remarked on that stage, but discarded 
it as being likely to have been caused by the emission from the accretion disc in the system 
becoming the dominant emission source. However, 
firstly, we find
this behaviour also likely to be present in [O{\sc iii}], that is,~in an emission 
that is exclusive to the shell. Secondly, especially in the later time ranges 
($\Delta t >$ 30 yr), the nebular remnant of those novae is already spatially 
resolved. 
For example, the most distant novae in our sample that can be suspected to
show this behaviour are CP Lac and V533 Her, with $d =$ 1.13 kpc and
1.17 kpc, respectively. 
Even assuming a comparatively
slow expansion of 500 km s$^{-1}$, their shells will have projected diameters of
about 5 arcsec. 
Thus, a contamination by emission from the accretion disc appears
unlikely. 
Finally, the ancient nova data are consistent with the idea that there is
another break in the decline law at late stages, with the slope being 
significantly diminished with respect to the main decline.

Taking all the above mentioned uncertainties into account, we suggest that
the luminosity evolution of all groups can be cautiously described as
experiencing an initial gentle decline, followed by a main steeper one,
which, at late stages, again returns to a much softer slope. Because of the
ambiguity in defining the time ranges of a specific phase, we do not
regard the differences in the times of transition from one phase to another
for different groups as significant as far as it concerns a specific
emission line.

\subsection{The main decline\label{maindecline_sec}}

We start our exploration of the physics behind the luminosity evolution
with the main decline because the initial stage can be best described
by recognising the differences 
to the former.

From Table \ref{fitpar_tab} and Fig.~\ref{fits_fig},
we see that for H$\alpha$, all groups present very similar slopes.
For the light curve types, we find a weighted mean of
$\bar{a}_\mathrm{LCT,H\alpha} = -2.46(05)$ and for the speed classes 
$\bar{a}_\mathrm{SC,H\alpha} = -2.49(07)$. For the [O{\sc iii}]
emission, the scatter is larger. Taking the weighted mean
of all light curve types yields $\bar{a}_\mathrm{LCT,[O\,III]} = -3.20(12)$. 
However, we note that there are two groups that deviate significantly 
from this value. 
In the O types, this might be due to the main decline 
being less well-defined as for the other groups, which is due to a lack of 
data in the comparatively large time range of 
$\log \Delta t \sim$0.5 .. $\sim$1.7 ($\sim$3 .. $\sim$50 yr). 
In the P class, the slope appears to be better defined, but still 
is strongly determined by an isolated point at very late stages. 
An absence of data at $\log \Delta t \sim$1.2 .. $\sim$1.7 
($\sim$16 .. $\sim$50 yr) allows for the potential
presence of a main decline plus late decline combination as found in
the S type, which would result in a steeper main decline.
Excluding these two groups from the averaging, the remaining three
yield $\bar{a}_\mathrm{LCT,[O\,III]} = -3.61(15)$, similar to
that for the speed classes, $\bar{a}_\mathrm{SC,[O\,III]} = -3.66(18)$.

As the referee has kindly pointed out on the first version of this article,
the closeness of the main decline slopes to a value of $-$3.0, suggesting that 
these largely reflect the volume change of the expanding shell.
The intensity $I_{\lambda}$ of an emission line at wavelength $\lambda$ 
along the path $dr$ depends on the ion and electron number densities 
$n_\mathrm{ion}$ and $n_e$ as
\begin{equation}
I_{\lambda} = \int n_\mathrm{ion}~n_e~\epsilon (\lambda, T_e)~dr ~,
\label{int_eq}
\end{equation}
where the emissivity $\epsilon (\lambda, T_e)$ is a function of $\lambda$ and
the temperature $T_e$, and includes the specifics of the atomic transition
probabilities. If we assume a homogeneous distribution of particles
and temperature, then the total flux integrated over an emission line that is
received from a shell of thickness $\Delta R$ is proportional to 
$n_\mathrm{ion}$, $n_e$, $\epsilon$ and the enclosed volume of emitting 
material $\Delta V \propto \Delta R^3$. There is growing evidence that the 
kinematics of the shell in the free expansion phase, without braking, can be 
best described as a ballistic expansion with the radius $R_s$ of the shell
increasing linearly with time $t$ \citep{masonetal18-1}. The geometry then
will be a self-similar function of time, so that $\Delta R(t) / R_s(t) =$ 
constant, and $\Delta V(t) \propto t^3$. At this point, also all material will 
be optically thin, so that the number of particles that contribute to the 
emission is constant. Thus, the densities from Eq.~\ref{int_eq} are both 
proportional to $t^{-3}$ and the total balance of these three parameters 
$n_\mathrm{ion}$, $n_e$ and $\Delta V$ yields a flux $F$ that is proportional 
to $t^{-3}$. Since this is very close to what is observed in 
Fig.~\ref{fits_fig}, the change in temperature $T_e$ appears to play only a 
minor role
in the luminosity evolution or it is cancelled out by other time dependent
factors that we have ignored here. Lastly (but not least), we point out that our 
assumption of a homogeneous distribution of densities and temperatures within 
the shell represents a gross simplification, as, for example,~\citet{williamsr13-2} 
suggests that nova shells present a clumpy structure already in the
initial stages of the eruption, with each clump having a specific density
and temperature distribution.

\subsection{The early stage\label{earlystage_sec}}

The luminosity behaviour up to about 100--200 d after maximum is described by 
an approximately constant slope. A few groups appear not to follow that 
tendency. However, in the S light curve type and the Very Fast speed classes, 
the comparatively steep initial decline ($a = -1.13(13)$) in the H$\alpha$ 
data is mainly determined by a single, very early data point of V1500 Cyg, 
while in the P and Fast classes this region is simply not covered by the 
available data. We note that in all these cases, the [O{\sc iii}] behaviour is
consistent with a very shallow decline or even constancy.

During this stage, the nova shell is still partly optically thick and
continues to be energised by the strong radiation of the eruption heated
white dwarf until the end of the nuclear burning, super-soft phase
\citep[e.g.][]{cunninghametal15-1}. As the shell expands, a steadily
larger growing part will become optically thin and, thus, contribute to the 
line emission. As the principal difference to the main decline we
thus identify that the number of particles, that is,~the emitting mass, is not 
constant, but increases up to the point where the complete shell has become
optically thin. Judging from Fig.~\ref{lumevol_fig}, this apparently
approximately cancels out the increase in volume $\Delta V$.

The initial shell luminosity is a measure for the energy released by the 
eruption. For comparison of the different groups, we take the mean of the
data points corresponding to this stage also in the cases where the slope
is not negligible. Taking into account the sparsity of the data and that
we are dealing with comparatively shallow slopes, this appears preferable
to simply taking the data point with the smallest $\Delta t$. These averages
are included, in addition to the actual linear fit, in Table \ref{fitpar_tab}.
Assuming that both the H$\alpha$ and the [O{\sc iii}] emission reflect the
differences between the individual groups in the same way, we find a sequence 
from higher to lower initial luminosities of O, S, D, J, P for the light curve types. For the speed classes, the behaviour of the H$\alpha$ emission 
faithfully reflects the decline rate $t_3$, with the Very Fast novae 
presenting the 
highest luminosity and the Slow and Very Slow novae the lowest. While this
stage is not covered in the Fast novae, the later data points at least do not
contradict a corresponding placement of that group within that sequence. 
However, for [O{\sc iii}], this tendency is not confirmed, with the Moderately
Fast novae showing the highest luminosities and the Fast novae the lowest. 
We  return to the differences between the speed class and the light
curve type behaviour in Section \ref{lctvssc_subsec}.

We note that for the light curve types, the end point in time of the initial 
stage follows the same sequence, in that the O and S groups are the first in
starting the main decline, while the J novae are the last. The only
exception here is the P class, which has a lower initial luminosity than
J, but ends this phase slightly earlier. As for the speed classes, other
than the initial luminosities, the sequence of the end of the initial phase
does not present any differences between H$\alpha$ and [O{\sc iii}] and
follows $t_3$. Still, as pointed out above, these breaking
points in the decline laws are prone to even larger uncertainties than the 
other parameters and should be interpreted with caution.

With that in mind, we consider how these findings can be related to physical properties.
As already mentioned, the initial luminosity corresponds to the energy of the
eruption. The end point of the initial stage could be interpreted as 
being defined by
a combination of the velocity of the ejecta and its mass, in
that less mass at higher velocities will become faster optically thin than a 
larger amount of material that has a lower velocity. The
consistency especially for the light curve types suggests that these
parameters are related, in that more luminous eruptions eject less material
and at higher velocities than less luminous ones. This is, in rough terms, 
indeed consistent with the models of \citet{yaronetal05-1}. However,
the underlying physical parameters of the binary remain ambiguous, because a 
combination of white dwarf masses and temperatures, as well as the accretion 
rate of the pre-nova, produces an overlapping parameter space of above 
eruption parameters and, thus, the observed behaviour cannot be tied to a
specific physical property.

\subsection{The late stage\label{latestage_sec}}

The ejecta  must necessarily interact
with the interstellar medium (ISM). With the works by \citet{duerbeck87-4} and
\citet{santamariaetal20-1}, there are two conflicting observational studies 
about the
braking effect of the ISM on the shell material during the first decades,
with the former finding that the braking coefficient is a function of the
initial ejection velocity, while the latter, in part for the same novae,
did not detect any measurable braking at all. In any case, at some point,
enough ISM material would have been swept up by the expanding ejecta to
enforce a significant braking \citep{oort46-1}. Supernova shells are known
to experience various stages of braking \citep[e.g.][chapter 4.9]{padmanabhan01-1}
and while the involved energies and masses are very different, the principal
behaviour should be similar.

In such a braking phase, the kinetic energy of the particles will
be transformed into radiation and the density in the affected regions will 
again increase. Thus, it is expected that the luminosity decline will
be softened with respect to the main decline phase. As pointed out above,
such behaviour is unambiguously only observed in our data for the S light
curve types and the Very Fast novae, but the data of the other groups are
still consistent with the existence of such a phase, that could start at
a later stage. We note that this also agrees well with the idea
that the S light curve types or the Very Fast novae eject less material
than the other groups (Section \ref{earlystage_sec}). 

The data of the ancient nova shells at $2.0 < \log \Delta t~[\mathrm{yr}] < 3.0$ 
certainly support the idea of a softened decline. Still, we have to keep
in mind that we are dealing with a handful of data points only, with the 
possibility
that these nova shells are only observed because of certain special conditions
in the surrounding ISM in this small number of objects. In fact, the large 
scatter is indicative of that at least not all these systems might be 
representative for the general behaviour. In particular, the shells of 
IGR J17014$-$4306 and of CRTS J054558.3+022106, and perhaps also of V1315 Aql
and BZ Cam, appear too bright by at least two to three orders of magnitude. We 
note that the uncertainty in the distance value for CRTS J054558.3+022106 
cannot account for such a large difference in luminosity. 

The most likely scenario for the brighter luminosities is the existence of 
higher density material in the vicinity of these novae prior to the eruption. 
One possibility is that this is due to the presence of previously ejected 
material, and that these could be comparatively `old' CVs that have 
experienced a large number of nova eruptions, perhaps with shorter recurrence 
times than other novae. We note that the data points for the recurrent novae
indeed appear to indicate brighter shells than for classical novae. If
we follow the argument of \citet{schaeferetal10-1} that the data point of 
T Pyx belongs to the shell from the 1866 eruption, it would be displaced
to $\log \Delta t = 2.1$, strengthening the similarity to the IGR J17014$-$4306
and CRTS J054558.3+022106 data. Another possibility is that the latter
two are actually comparatively young novae and are embedded in the remnants
of a planetary nebula \citep{wessonetal08-3,rodriguez-giletal10-1,%
jonesdetal19-1}. Detailed abundance analyses should be able to distinguish
between the two scenarios \citep{wessonetal18-2}.

The data on the other three ancient novae agree well with any of the light
curve types or speed classes and if we may draw any conclusions
from a mere three data points, they appear to indicate that the luminosity
decline is only slowed down, but not completely stopped. Corresponding 
luminosity data on the oldest such nova, Z Cam, would certainly be helpful 
in testing this trend. In any case, since the rate of discovery of ancient
novae has significantly increased over the last few years, there is hope that
more data can be added in the not-too-distant future.

\subsection{Light curve type versus speed class\label{lctvssc_subsec}}

One of the goals of this study is to investigate whether there is a better
way of grouping novae with respect to a certain parameter than the speed
class. At first glance, the latter appears as a reasonable choice because
the rate of decline habitually has been thought to be related to the
energy of the eruption and the velocity of the ejection \citep{shara81-2},
giving rise to the absolute maximum magnitude versus rate of decline
relation \citep[MMRD,][]{mclaughlin45-2}. 
However, during the last decade,
the discovery of extragalactic novae that do not fit the MMRD at all
\citep{kasliwaletal11-2,sharaetal17-3} and the revised determination 
of absolute magnitudes of novae based on DR2 distances \citep{schaefer18-2}
placed severe doubts on the validity of this relation. 
Based on a small sample of novae with well-determined distances,
\citet{selvelli+gilmozzi19-1} present a new parametrisation of the
MMRD, but their fit to the data (their figure 1) still shows possibly 
systematic residuals.

The grouping according to light curve type is motivated by the assumption
that many of the features seen in the broad-band photometry could be
present in the shell emission as well. Indeed, we find that the dust dips
that define the D light curve type are also detected in the H$\alpha$ and
[O{\sc iii}] emission. Thus, taking into account the broad-band light curve
allows us to exclude the points that represent a systematic deviation
from the general decline law (see also our remarks on V2214 Oph in
Section \ref{evol_sec}). With respect to the other classes, the
corresponding features either occur at too small time scales and flux scales 
to be unambiguously detected in our data (oscillations and jitter) or
fall into the gaps due to undersampling (plateaus, flat-tops and cusps,
the latter not being represented at all in our sample).

Comparing the quality of the fits for the two groupings, we find that
while the scatter is, on average, smaller for the light curve type, the
differences are not substantial (Table \ref{fitpar_tab}) and, as shown in 
Section \ref{gendesc_subsec}, also the derived slopes of the main decline 
are identical within the errors. However, a close look at Fig.~\ref{lumevol_fig}
shows that the light curve groups overall present less systematic scatter.
This is especially evident in the case of the [O{\sc iii}] emission of 
the Very Fast speed class that includes the S type novae CP Lac and V1500 Cyg
as well as the O types V603 Aql and GK Per, which present a significant
displacement when placed together in the speed class, but fit well when 
separated by light curve type. However, even in the light curve groups,
we find a number of systems that do show systematically different behaviour
(e.g.~PW Vul and HR Del in the P group), indicating that the light
curve types still harbour a diversity that is significant for the luminosity
evolution.

\subsection{{\rm H$\alpha$} versus {\rm{[O{\sc iii}]}}\label{hvso_subsec}}

The two emission lines investigated in this study represent two different
types of atomic transitions, allowed and forbidden. As such, they
require different density conditions and thus track different regions in 
the shell material. Comparing the luminosity evolution of the two lines
we confirm the general result of \citetalias{downesetal01-2} in that
the [O{\sc iii}] emission appears later and declines more rapidly than
H$\alpha$. 
In the lower plot of Fig.~\ref{fits_fig}, we compare the slopes of the
main decline stage for the different groups and the two emission lines.
We find that the H$\alpha$ emission line consistently have shallower slopes
than the `theoretical' value of -3.0 (Section \ref{maindecline_sec}),
while most [O{\sc iii}] groups present a steeper decline. 
Both are also to be expected, because firstly, as the expanding
shell evolves from high to low density, the conditions for forbidden
transitions will develop later than those for allowed ones. Secondly,
in the higher density hydrogen-emitting regions, emitted photons 
(in this case mainly Ly$\alpha$) can be absorbed, thus re-heating
the material, while the low density in the [O{\sc iii}] regions yields 
more efficient cooling \citep{becketal90-4}.

There are a few exceptions to this behaviour. Defining the ratio of the
main slopes as discussed in Section \ref{gendesc_subsec} as
\begin{equation}
f = \frac{a_\mathrm{H\alpha}}{a_\mathrm{[O\,III]}}~,
\end{equation}
we find $f = 0.68(04)$ for the average of the D, J, S light curve types 
(0.68(05) for the speed classes), while the P types have $f = 0.98(21)$ and 
V838 Her has $f = 1.31(33)$, that is,~the H$\alpha$ and the [O{\sc iii}]
luminosities decline at roughly the same rate. 
The general impression from the recurrent novae, and especially from
the V3890 Sgr data, is that they follow a similar trend of 
$f \approx 1$.
The result for the
P novae includes a large uncertainty and still agrees with the other
light curve types within 1.5$\sigma$, so that we have to be careful not
to fall into the trap of overinterpretation.
However, we remark that
\citetalias{stropeetal10-1} speculate that this class harbours yet
unrecognised recurrent novae since a plateau phase caused by
a very luminous accretion disc appears to be a typical feature of
recurrent novae light curves \citep[e.g.][]{hachisuetal00-1,hachisuetal03-1}.
Both the comparatively large scatter and the slope ratio could thus 
possibly be explained by this class containing a mixture of classical
and recurrent novae. V838 Her also has been flagged as a strong
candidate for a recurrent nova for a number of reasons 
\citep{pagnotta+schaefer14-1}. 
Still, we note that V838 Her actually does not seem to fit well into any 
category. Also, the data on the recurrent novae are not uniformly distributed,
so that we do not find sufficient grounds for investigating a possible 
diversity among these systems.
\citetalias{downesetal01-2} distinguish between 
recurrent novae that 
contain
an evolved secondary star and those where the
donor is still close to the main-sequence, finding, indeed, different slopes.
In our diminished sample, RS Oph and V3890 Sgr are systems with evolved
secondaries, comparatively long orbital periods of 456 d and 520 d, respectively
\citep{fekeletal00-1,schaefer09-2}, and wind accretion, while T Pyx has a
short orbital period of 1.8 h \citep{uthasetal10-1} and accretion via
Roche-lobe overflow.  
Still, the data suggest that a common property might be that they end up
with brighter shells than classical novae and we could speculate that this
is due to interaction with the nova shell remnants from previous eruptions.

Finally, we note that also the O light curve type presents an [O{\sc iii}]
decline with a shallower slope $> -3.0$ than the P, J and S groups. However,
the slope is mainly defined by the behaviour of GK Per, which has, for classical novae, an unusually long orbital period of 2 d 
\citep{cramptonetal86-1} and, additionally, is likely to be embedded in the remnants 
of a planetary nebula \citep{bodeetal87-3,harveyetal16-2}. It is, therefore,
likely that it is not representative of other novae of this class. In fact, the,
admittedly sparse and unevenly distributed, data of V603 Aql suggest a
steeper slope.

\section{Summary and conclusions}

In this paper, we present a re-analysis of the H$\alpha$ and [O{\sc iii}] flux
data of nova shells collected by \citet{downesetal01-2}, using the interstellar 
reddening values from \citet{oezdoenmezetal16-2} and the distances from the Gaia 
DR2 archive, corrected for the influence of the Galactic potential by 
\citet{bailer-jonesetal18-4}. 
With this aim, we carefully revised the identifications of the novae in our
sample with Gaia sources and analysed the validity of the distance values.
We used two different criteria to group the 
individual data points, one being the speed class \citep{payne-gaposchkin64-1} 
the other the light curve type \citep{stropeetal10-1}.
        
We find that grouping according to light curve type is advantageous 
compared to the speed class because it yields less systematic scatter and
allows us to more easily to identify data that do not represent the
intrinsic shell luminosity (e.g.~dust dips). The main weakness of the 
light curve type grouping is that the necessity
for a well-documented photometric light curve limits the number of
systems that it can be applied to. Spreading the data over (in principle) 
seven groups further enhances the effect of undersampling.
This yields distributions that either present large gaps or are dominated by a 
single object (the O group is a good example for both).
Overall, however, the behaviour appears to be consistent
for all groups.

In general terms, the evolution can be divided into three stages.
In the logarithmic representation, an initial soft decline or 
constant behaviour is followed by a main decline which, at late
stages, potentially transforms again into a much more gradual
decline or constancy. 
We tentatively ascribe the physical processes behind the three stages as:
the initial expansion of the shell, transitioning from an optically thick to
a fully optically thin configuration, where all ejected material contributes 
to the line emission; the
subsequent free expansion of the optically thin material, with the 
decline in luminosity being mainly determined by the increasing volume and
the decreasing densities; and, finally, the interaction with the surrounding
interstellar medium. Additional data on `ancient' nova shells with ages
$>$130 yr agree well with the latter. We also find that at least two of
these ancient novae, IGR J17014$-$4306 and CRTS J054558.3+022106, have
significantly brighter shells, which is possibly due to the existence of
denser material prior to the nova eruption, perhaps in the form of
the remnants of previous nova eruptions or of planetary nebulae.

Confirming the results from \citet{downesetal01-2},
the [O{\sc iii}] emission appears later \citep[$\sim$20 d after eruption,
see also e.g.][]{ederocliteetal06-1} and
declines significantly faster than H$\alpha$.
The only exceptions from this
behaviour are demonstrated (possibly) by the recurrent novae, the P light curve type systems, 
and V838 Her, the latter two being suspected to also be related to
recurrent novae.

Almost every group includes a small number of objects which, while 
being close to the bulk of the data, present a systematically different
slope. In addition, there are also a number of isolated cases (novae with
only one or two data points) that appear to be further detached
from the general distribution. It might thus be advisable to discuss
the behaviour of individual systems and look for a common denominator
of similar declines instead of the approach employed by \citet{downesetal01-2}, which is based on a grouping that already presumes the validity of 
a specific common denominator, motivated by the lack of
sufficient data for the majority of novae. In a forthcoming paper, we
will present new luminosity data and an investigation on a case-by-case
basis.

\begin{acknowledgements}
We thank the anonymous referee for helpful comments that managed to both
expand the scope of this article and tighten the science case at the same
time.

CT, NV and MV acknowledge financial support from Conicyt-Fondecyt grant
No.~1170566.

This work has made use of data from the European Space Agency (ESA) mission
{\it Gaia} (\url{https://www.cosmos.esa.int/gaia}), processed by the {\it Gaia}
Data Processing and Analysis Consortium (DPAC,
\url{https://www.cosmos.esa.int/web/gaia/dpac/consortium}). Funding for the DPAC
has been provided by national institutions, in particular the institutions
participating in the {\it Gaia} Multilateral Agreement.
\end{acknowledgements}


\begin{appendix} 

\section{Nova parameters\label{novapar_sec}}

\longtab[1]{
\begin{longtable}{llllllrllll}
\caption{Coordinates and DR2 parallaxes on all novae that are listed in 
\citetalias{downesetal01-2} with [O{\sc iii}] or H$\alpha$ fluxes, as 
well as for the ancient novae (indicated as AN). In the interest of space, the 
designations of the three ancient novae from surveys have been abbreviated. 
They are given in full in Section \ref{an_sec}. The coordinates in 
columns 3 and 4 were taken from the \citet{downesetal05-1} catalogue or 
above mentioned surveys, unless noted otherwise.  $|\Delta_\alpha|$ and 
$|\Delta_\delta|$ represent the absolute difference of these coordinates
with those of the DR2 source identified with the nova. $\bar{\omega}$, 
$\sigma_{\bar{\omega}}$ and $f_{\bar{\omega}}$ give the DR2 parallax, its 
uncertainty and the ratio of the two, $f_{\bar{\omega}} = 
\sigma_{\bar{\omega}} / \bar{\omega}$, respectively. The LCT column states the 
light curve type from \citet{stropeetal10-1}. Here, we  opted to place the 
recurrent novae in an extra category, marked `RN'. Finally, in the last column, 
we mark those objects with an asterisk where we provide an additional commentary.
\label{allD01_tab}}\\
\hline
\hline\noalign{\smallskip}
Object & DR2 source & $\alpha$ (J2000.0) & $\delta$ (J2000.0) & $|\Delta_\alpha|$ & $|\Delta_\delta|$ 
& $\bar{\omega}$ & $\sigma_{\bar{\omega}}$ & $f_{\bar{\omega}}$ & LCT & Note \\
\hline\noalign{\smallskip}
\endfirsthead
\caption{continued.}\\
\hline\noalign{\smallskip}
Object & DR2 source & $\alpha$ (J2000.0) & $\delta$ (J2000.0) & $|\Delta_\alpha|$ & $|\Delta_\delta|$ 
& $\bar{\omega}$ & $\sigma_{\bar{\omega}}$ & $f_{\bar{\omega}}$ & LCT & Note \\
\hline\noalign{\smallskip}
\endhead
\hline\noalign{\smallskip}
\endfoot
\hline
\endlastfoot
OS And    & 1942264441241366144 & 23:12:05.95 & $+$47:28:19.6 & 0.25 & 0.12 & 0.14    & 0.14 & 1.00 & D  &   \\
DO Aql    & 4208116120905286528 & 19:31:25.88 & $-$06:25:38.8 & --   & --   & 0.77    & 0.30 & 0.39 & F  & * \\
V603 Aql  & 4266547566124966912 & 18:48:54.64 & $+$00:35:02.9 & 0.13 & 0.19 & 3.19    & 0.07 & 0.02 & O  &   \\
V1315 Aql & 4313192491505026944 & 19:13:54.54 & $+$12:18:03.5 & 0.17 & 0.36 & 2.23    & 0.03 & 0.01 & AN &   \\
V1370 Aql & 4288898099224201856 & 19:23:21.24 & $+$02:29:26.3 & 0.02 & 0.10 & 0.34    & 0.19 & 0.55 & D  & * \\
V1419 Aql & 4267751638733390592 & 19:13:06.80 & $+$01:34:23.3 & 0.14 & 0.06 & 5.75    & 1.87 & 0.33 & D  &   \\
V1425 Aql & --                  & 19:05:26.63 & $-$01:42:03.3 & --   & --   & --      & --   & --   & S  & * \\
V341 Ara  & 5818105674339588096 & 16:57:41.51 & $-$63:12:38.4 & 1.64 & 1.34 & 6.40    & 0.08 & 0.01 & AN &   \\
T Aur     & 3446266197646225536 & 05:31:59.12 & $+$30:26:45.1 & 0.01 & 0.16 & 1.14    & 0.05 & 0.04 & D  &   \\
Z Cam     & 1123169888190445568 & 08:25:13.18 & $+$73:06:39.2 & 0.20 & 0.22 & 4.44    & 0.04 & 0.01 & AN &   \\
BZ Cam    & 1112772429499375488 & 06:29:34.00 & $+$71:04:34.3 & 0.71 & 1.65 & 2.69    & 0.04 & 0.01 & AN &   \\
AT Cnc    & 679528804789642240  & 08:28:36.92 & $+$25:20:03.0 & 0.04 & 0.09 & 2.20    & 0.05 & 0.02 & AN &   \\
V365 Car  & 5338717294658032128 & 11:03:15.82 & $-$58:27:26.1 & 0.61 & 0.07 & 0.28    & 0.13 & 0.47 & -- &   \\
V705 Cas  & 1998676229637930496 & 23:41:47.20 & $+$57:30:59.8 & 0.44 & 0.99 & 0.48    & 0.09 & 0.18 & D  &   \\
V723 Cas  & 411250132279713280  & 01:05:05.37 & $+$54:00:40.5 & 0.24 & 0.27 & 0.13    & 0.05 & 0.36 & J  &   \\
V812 Cen  & 6063092872344818048 & 13:13:54.32 & $-$57:40:44.4 & 2.07 & 0.35 & $-$0.66 & 1.24 & 1.87 & -- & * \\
V842 Cen  & 5891405647833287296 & 14:35:52.55 & $-$57:37:35.3 & 0.13 & 0.19 & 0.73    & 0.05 & 0.07 & D  &   \\
V868 Cen  & 5864691672765615616 & 13:50:10.70 & $-$63:08:52.0 & --   & --   & 1.15    & 0.88 & 0.77 & J  & * \\
V888 Cen  & 6056093557454448512 & 13:02:31.86 & $-$60:11:38.3 & --   & --   & 0.34    & 0.11 & 0.34 & O  & * \\
IV Cep    & 2005311404366428544 & 22:04:36.91 & $+$53:30:23.6 & --   & --   & 0.11    & 0.05 & 0.51 & -- & * \\
BY Cir    & 5874108818084122624 & 14:44:53.46 & $-$63:53:55.8 & 0.02 & 0.17 & 0.30    & 0.17 & 0.56 & P  &   \\
V394 CrA  & 4035768120009351936 & 18:00:25.97 & $-$39:00:35.1 & 0.84 & 0.63 & 0.22    & 0.26 & 1.17 & RN &   \\
CP Cru    & 6057768830885584896 & 12:10:31.33 & $-$61:45:09.7 & 0.31 & 0.15 & 0.67    & 0.40 & 0.60 & -- &   \\
Q Cyg     & 1966874711229398656 & 21:41:43.93 & $+$42:50:29.1 & 0.02 & 0.09 & 0.73    & 0.02 & 0.03 & -- &   \\
V450 Cyg  & 1869766290946348160 & 20:58:47.39 & $+$35:56:27.9 & 1.16 & 0.07 & $-$0.33 & 0.24 & 0.74 & -- & * \\
V476 Cyg  & 2089624258071404544 & 19:58:24.50 & $+$53:37:06.8 & --   & --   & 1.52    & 0.17 & 0.11 & D  & * \\
V1363 Cyg & 2058291887543939968 & 20:06:11.53 & $+$33:42:37.7 & 0.12 & 0.14 & 0.56    & 0.05 & 0.08 & AN &   \\
V1500 Cyg & 2165295912482637312 & 21:11:36.51 & $+$48:09:02.8 & 0.91 & 0.93 & 0.78    & 0.19 & 0.24 & S  &   \\
V1819 Cyg & 2059638205176483712 & 19:54:37.57 & $+$35:42:16.0 & 0.37 & 0.59 & 0.11    & 0.55 & 4.95 & J  &   \\
V1974 Cyg & 2181756563616247552 & 20:30:31.67 & $+$52:37:50.8 & 0.29 & 0.06 & 0.62    & 0.07 & 0.11 & P  &   \\
HR Del    & 1813953083546374144 & 20:42:20.35 & $+$19:09:39.3 & 0.08 & 0.14 & 1.04    & 0.03 & 0.03 & J  &   \\
DQ Her    & 2116226254706461568 & 18:07:30.26 & $+$45:51:32.1 & 0.15 & 0.66 & 2.00    & 0.02 & 0.01 & D  &   \\
V446 Her  & 4506083222297339008 & 18:57:21.61 & $+$13:14:28.6 & 0.32 & 0.62 & 0.74    & 0.07 & 0.10 & S  &   \\
V533 Her  & 2113091615775375232 & 18:14:20.48 & $+$41:51:22.1 & 0.08 & 0.01 & 0.83    & 0.03 & 0.04 & S  &   \\
V827 Her  & --                  & 18:43:42.60 & $+$15:19:19.0 & --   & --   & --      & --   & --   & S  &   \\
V838 Her  & 4504548029183559552 & 18:46:31.46 & $+$12:14:02.0 & 0.13 & 0.00 & 0.46    & 0.71 & 1.53 & P  & * \\
CP Lac    & 2006109065688505472 & 22:15:41.09 & $+$55:37:01.3 & 0.02 & 0.03 & 0.86    & 0.04 & 0.05 & S  &   \\
DK Lac    & 2002440098459791744 & 22:49:46.97 & $+$53:17:20.0 & 0.63 & 0.23 & 0.40    & 0.07 & 0.17 & J  &   \\
HY Lup    & 5898781618479432064 & 14:31:50.21 & $-$51:10:32.3 & 0.67 & 0.25 & 0.15    & 0.76 & 5.24 & -- &   \\
BT Mon    & 3106991818813980416 & 06:43:47.20 & $-$02:01:14.5 & 0.59 & 0.61 & 0.68    & 0.05 & 0.07 & F  & * \\
GQ Mus    & 5236081560713688448 & 11:52:02.48 & $-$67:12:20.1 & 0.00 & 0.87 & 0.47    & 0.22 & 0.47 & -- & * \\
RS Oph    & 4174878674679897344 & 17:50:13.16 & $-$06:42:28.5 & 0.01 & 0.07 & 0.44    & 0.05 & 0.12 & RN &   \\
V841 Oph  & 4333061392472253440 & 16:59:30.37 & $-$12:53:27.2 & 0.00 & 0.02 & 1.21    & 0.03 & 0.02 & -- &   \\
V972 Oph  & 4060410267160342272 & 17:34:44.48 & $-$28:10:35.5 & 0.39 & 0.32 & 0.91    & 0.24 & 0.26 & -- & * \\
V2104 Oph & 4495453590539471488 & 18:03:24.99 & $+$11:47:57.1 & 0.00 & 0.06 & $-$0.19 & 0.84 & 4.51 & -- & * \\
V2214 Oph & 6028898202535507328 & 17:12:01.58 & $-$29:37:33.3 & 1.72 & 0.50 & 2.16    & 0.61 & 0.28 & S  & * \\
V2264 Oph & --                  & 17:20:20.83 & $-$26:46:26.3 & --   & --   & --      & --   & --   & S  & * \\
GK Per    & 238540495056450048  & 03:31:12.01 & $+$43:54:15.4 & 0.15 & 0.20 & 2.26    & 0.04 & 0.02 & O  &   \\
V400 Per  & 435763400421947264  & 03:07:38.24 & $+$47:07:40.0 & 0.37 & 0.48 & $-$0.39 & 0.63 & 1.59 & -- &   \\
RR Pic    & 5477422099543150592 & 06:35:36.07 & $-$62:38:24.3 & 0.01 & 0.05 & 1.96    & 0.03 & 0.02 & J  &   \\
CP Pup    & 5544760551021856000 & 08:11:46.06 & $-$35:21:05.0 & 0.02 & 0.05 & 1.23    & 0.02 & 0.02 & P  &   \\
V351 Pup  & 5544772684308712192 & 08:11:38.40 & $-$35:07:30.5 & 0.12 & 0.23 & 0.09    & 0.51 & 5.91 & P  &   \\
T Pyx     & 5628258258606112768 & 09:04:41.51 & $-$32:22:47.6 & 0.10 & 0.10 & 0.31    & 0.04 & 0.14 & RN &   \\
V3888 Sgr & --                  & 17:48:40.55 & $-$18:45:36.8 & --   & --   & --      & --   & --   & -- &   \\
V3890 Sgr & 4077352126434336640 & 18:30:43.29 & $-$24:01:08.9 & 0.10 & 0.08 & 0.19    & 0.09 & 0.48 & RN &   \\
V4157 Sgr & --                  & 18:09:34.90 & $-$25:51:58.2 & --   & --   & --      & --   & --   & -- & * \\
V4160 Sgr & --                  & 18:14:13.83 & $-$32:12:28.5 & --   & --   & --      & --   & --   & S  & * \\
V4169 Sgr & 4051419667958619776 & 18:23:26.94 & $-$28:21:59.7 & 0.16 & 0.08 & $-$0.06 & 0.48 & 8.06 & S  & * \\
V4171 Sgr & --                  & 18:23:41.34 & $-$22:59:28.7 & --   & --   & --      & --   & --   & -- & * \\
V4361 Sgr & --                  & 18:23:42.46 & $-$18:07:14.7 & --   & --   & --      & --   & --   & -- & * \\
V4444 Sgr & --                  & 18:07:36.22 & $-$27:20:13.5 & --   & --   & --      & --   & --   & S  & * \\
V4633 Sgr & --                  & 18:21:40.47 & $-$27:31:38.0 & --   & --   & --      & --   & --   & P  & * \\
V4642 Sgr & --                  & 17:55:09.84 & $-$19:46:01.0 & --   & --   & --      & --   & --   & -- & * \\
U Sco     & 6246188565119443072 & 16:22:30.81 & $-$17:52:44.1 & 0.47 & 0.82 & $-$0.35 & 0.21 & 0.61 & RN &   \\
V745 Sco  & 4043221606260719360 & 17:55:22.25 & $-$33:14:59.5 & 0.39 & 0.80 & $-$0.67 & 0.31 & 0.46 & RN &   \\
V960 Sco  & --                  & 17:56:34.14 & $-$31:49:36.3 & --   & --   & --      & --   & --   & -- & * \\
V977 Sco  & --                  & 17:51:50.35 & $-$32:32:00.1 & --   & --   & --      & --   & --   & -- & * \\
V992 Sco  & 5965441877425525248 & 17:07:17.44 & $-$43:15:21.6 & 0.21 & 0.37 & 0.40    & 0.15 & 0.38 & D  &   \\
V1141 Sco & --                  & 17:54:11.20 & $-$30:02:52.0 & --   & --   & --      & --   & --   & -- & * \\
V1142 Sco & --                  & 17:55:24.99 & $-$31:01:41.5 & --   & --   & --      & --   & --   & -- & * \\
FV Sct    & 4104903001614210560 & 18:34:51.68 & $-$12:55:26.4 & --   & --   & 0.63    & 0.29 & 0.46 & -- & * \\
V373 Sct  & 4204056513527494784 & 18:55:26.71 & $-$07:43:05.5 & 1.71 & 0.70 & 0.29    & 0.40 & 1.37 & J  & * \\
V443 Sct  & --                  & 18:49:38.95 & $-$06:11:15.9 & --   & --   & --      & --   & --   & J  & * \\
V444 Sct  & --                  & 18:47:09.91 & $-$08:20:53.6 & --   & --   & --      & --   & --   & -- & * \\
X Ser     & 4358729079104357760 & 16:19:17.71 & $-$02:29:30.0 & 0.46 & 0.28 & 0.29    & 0.14 & 0.50 & -- &   \\
CT Ser    & 1192697922589844352 & 15:45:39.07 & $+$14:22:32.1 & 0.14 & 0.69 & 0.23    & 0.06 & 0.27 & -- &   \\
FH Ser    & 4276984993803967744 & 18:30:47.06 & $+$02:36:52.7 & 0.27 & 0.72 & 0.95    & 0.08 & 0.08 & D  & * \\
LW Ser    & --                  & 17:51:50.89 & $-$14:43:50.6 & --   & --   & --      & --   & --   & D  & * \\
XX Tau    & 3394342581362155520 & 05:19:24.46 & $+$16:43:00.7 & 0.08 & 0.27 & 0.39    & 0.33 & 0.86 & -- &   \\
RW UMi    & 1704994848488583552 & 16:47:54.78 & $+$77:02:12.2 & 0.54 & 0.18 & 0.47    & 0.19 & 0.41 & -- &   \\
LV Vul    & 2027963130624974080 & 19:48:00.70 & $+$27:10:19.5 & 0.30 & 0.05 & 1.10    & 0.13 & 0.12 & S  & * \\
NQ Vul    & 2017742684676480896 & 19:29:14.72 & $+$20:27:59.3 & 0.52 & 0.22 & 0.95    & 0.10 & 0.11 & D  &   \\
PW Vul    & 2025732152809113600 & 19:26:05.03 & $+$27:21:57.0 & 0.30 & 1.10 & 0.43    & 0.10 & 0.23 & J  &   \\
QU Vul    & 1860040595206017664 & 20:26:46.02 & $+$27:50:43.2 & 0.02 & 0.09 & 0.74    & 0.37 & 0.49 & P  &   \\
QV Vul    & 4520655771453037184 & 19:04:40.29 & $+$21:46:14.1 & 0.12 & 0.01 & 0.13    & 0.21 & 1.71 & D  &   \\
CRTS J05  & 3223064065896253056 & 05:45:58.26 & $+$02:21:06.2 & 0.03 & 0.05 & 1.54 & 0.96 & 0.62 & AN &  \\
IGR J17   & 5966150998003864960 & 17:01:28.15 & $-$43:06:12.3 & 0.03 & 0.16 & 0.96   & 0.05 & 0.05 & AN &   \\
IPHASX J21 & 2163877198882886656 & 21:02:05.82 & $+$47:10:18.0 & 0.17 & 0.00 & 1.35 & 0.04 & 0.03 & AN & \\
\end{longtable}
}

\begin{table*}
\setlength{\tabcolsep}{4pt}
\caption{Parameters of the individual novae: light curve type LCT 
\citepalias{stropeetal10-1}, speed class SC \citepalias{downesetal01-2},
DR2 distance $d_\mathrm{BJ}$ \citep{bailer-jonesetal18-4}, the interstellar
reddening $E(B-V)$ 
\citep{oezdoenmezetal18-1}, and,
for each emission line, the number of data points $n$ and the time range 
covered by the data $\Delta t_\mathrm{tot}$. The parentheses in the $n$ 
columns indicate the number of data points excluded from the fit. 
For other values in parentheses, see text, especially Section \ref{sample_sec}.
For the ancient novae (AN),
the reddening values have been taken from the Stilism website 
\citep{lallementetal19-1}.
} 
\label{novaprop_tab}
\centering
\begin{tabular}{llllllllll}
\hline
\hline\noalign{\smallskip}
Object & LCT & SC & $d_\mathrm{BJ}$ & $E(B-V)$ 
&  $n_\mathrm{H\alpha}$ & $\Delta t_\mathrm{tot,H\alpha}$  
& $n_\mathrm{[O\,III]}$ & $\Delta t_\mathrm{tot,[O\,III]}$  
& Notes\\
 & & & [$10^3$ pc] & [mag] & & [yr] & & [yr] & \\
\hline\noalign{\smallskip}
DO Aql     & F & S  & 1.5$^{+1.7}_{-0.6}$ & 0.13
& 1       &  74.9        & 1     &  74.9
& $E(B-V)$ from \citetalias{downesetal01-2} \\[0.08cm]
V603 Aql  & O & VF & 0.311$^{+0.007}_{-0.007}$ & 0.08(02)
& 3       & 72.2 .. 78.0 & 20    & 0.07 .. 59.8
& \\[0.08cm]
V1315 Aql & -- & -- & 0.443$^{+0.006}_{-0.006}$ & 0.14(07)
& 1       & 850$^{+350}_{-350}$ & -- & --
& AN \\[0.08cm]
V1370 Aql & D & F  & 2.9$^{+2.4}_{-1.1}$ & 0.6      
& 2       & 0.4 .. 9.5   & 1     & 0.4
& $E(B-V)$ from \citetalias{downesetal01-2}  \\[0.08cm]
V341 Ara  & -- & -- & 0.155$^{+0.003}_{-0.001}$ & 0.033(09)
& 1       & 800          & 1     & 800
& AN \\[0.08cm]
T Aur     & D & MF & 0.86$^{+0.04}_{-0.04}$ & 0.42(08)
& 1       & 94.0         & 1     & 87.2
& \\[0.08cm]
AT Cnc    & -- & -- & 0.449$^{+0.009}_{-0.010}$ & 0.028(10)
& 1       & 330$^{+135}_{-90}$ & -- & --
& AN \\[0.08cm]  
BZ Cam    & -- & -- & 0.368$^{+0.005}_{-0.005}$ & 0.05(04)
& 1       & 1630         & 1     & 1630 
& AN \\[0.08cm]
V705 Cas  & D & MF & 2.0$^{+0.4}_{-0.3}$ & 0.41(06)
& 1       & 0.19         & 1     & 5.0
&  \\[0.08cm]
V723 Cas  & J & S  & 4.9$^{+1.1}_{-0.8}$ & 0.45
& 3       & 0.1 .. 2.3   & 1     & 2.3
& \\[0.08cm]
V842 Cen  & D & MF & 1.3$^{+0.1}_{-0.1}$ & 0.55(05)
& 5       & 0.6 .. 13.7  & 5     & 0.6 .. 13.7
& \\[0.08cm]
V888 Cen  & O & F  & 2.9$^{+1.0}_{-0.9}$ & 0.34
& 1       & 3.1          & 2     & 2.8 .. 3.1 
& \\[0.08cm]
BY Cir    & P & F  & 3.1$^{+2.3}_{-1.1}$ & 0.13(06)
& 1       & 3.2          & 1     & 3.2
& \\[0.08cm]
V476 Cyg  & D & VF & 0.66$^{+0.09}_{-0.07}$ & 0.18(10)
& 2       & 42.2 .. 64.0 & 6 (3) & 0.08 .. 0.15
& \\[0.08cm]
V1500 Cyg & S & VF & 1.3$^{+0.5}_{-0.3}$ & 0.45(07)
& 16      & 0.01 .. 14.9 & 13    & 0.1 .. 9.0
& \\[0.08cm]
V1974 Cyg & P & F  & 1.6$^{+0.2}_{-0.2}$ & 0.26(03) 
& 6       & 0.5 .. 6.3   & 6     & 0.7 .. 6.3 
& \\[0.08cm]
HR Del    & J & VS & 0.93$^{+0.03}_{-0.03}$ & 0.17(02)
& 8       & 0.2 .. 33.0  & 16    & 0.2 .. 33.0
& \\[0.08cm]
DQ Her    & D & MF & 0.494$^{+0.005}_{-0.006}$ & 0.05(02)
& 8       & 0.03 .. 60.4 & 9 (3) & 0.44 .. 60.4
& \\[0.08cm]  
V446 Her  & S & F  & 1.3$^{+0.1}_{-0.1}$ & 0.37(04)
& 1       & 37.3         & 0     & --
& \\[0.08cm]
V533 Her  & S & F  & 1.17$^{+0.05}_{-0.04}$ & 0.03(02)
& 2       & 33.3 .. 35.1 & 1     & 15.2 
& \\[0.08cm]
V838 Her  & P & VF & (3.2$^{+3.1}_{-1.8}$) & 0.49
& 13      & 0.03 .. 1.5  & 10    & 0.05 .. 0.61
& $E(B-V)$ from \citetalias{downesetal01-2} \\[0.08cm]
CP Lac    & S & VF & 1.13$^{+0.05}_{-0.05}$ & 0.27(06)
& 11      & 0.04 .. 54.1 & 8     & 0.07 .. 43.3
& \\[0.08cm]
DK Lac    & J & F  & 2.3$^{+0.5}_{-0.3}$ & 0.22(06)
& 1       & 41.4         & 1     & 29.8
& \\[0.08cm]
BT Mon    & F & (S)  & 1.4$^{+0.1}_{-0.1}$ & 0.24(06)
& 1 (1)   & 42.2         & 1 (1) & 40.1
& no SC assigned by \citetalias{downesetal01-2} \\[0.08cm]
RS Oph    & (P) & (VF) & (2.1$^{+0.3}_{-0.2}$) & 0.73(10)
& 8       & 0.001 .. 0.029 & 5   & 0.07 .. 0.55
& RN \\[0.08cm]
V2214 Oph & (S) & MF & 0.6$^{+2.9}_{-0.2}$ & 0.73(10)
& 20 (20) & 0.07 .. 13.3 & 1 (1) & 13.3
& LCT is not S \\[0.08cm]
GK Per    & O & VF & 0.44$^{+0.01}_{-0.01}$ & 0.34(04)
& 2       & 83.6 .. 84.0 & 11    & 0.2 .. 84.0 
& \\[0.08cm]
RR Pic    & J & S  & 0.504$^{+0.008}_{-0.008}$ & 0.00(02)
& 1       & 72.8         & 18    & 0.7 .. 72.8
& \\[0.08cm]
CP Pup    & P & VF & 0.80$^{+0.01}_{-0.01}$ & 0.2
& 1       & 52.9         & 1     & 52.9
& \\[0.08cm]
T Pyx     & (P) & (MF) & 2.9$^{+0.4}_{-0.3}$ & 0.25(02)
& 1       & (31.3)         & 1     & (31.3)
& RN \\[0.08cm]
V3890 Sgr & (S) & (VF) & (4.4$^{+2.5}_{-1.3}$) & 0.90(30)
& 11      & 0.01 .. 1.94 & 4     & 0.18 .. 0.5
& RN \\[0.08cm]
V992 Sco  & D  & MF & 2.4$^{+1.5}_{-0.7}$ & 1.3(1)
& 25 (2)  & 0.02 .. 1.1  & 3     & 0.8 .. 1.1
& \\[0.08cm]
LV Vul    & S & F  & 0.9$^{+0.1}_{-0.1}$ & 0.57(05)
& 1       & 23.3         & 0     & --
& \\[0.08cm]
NQ Vul    & D & F  & 1.0$^{+0.1}_{-0.1}$ & 0.92(20)
& 2       & 14.7 .. 21.6 & 2     & 14.7 .. 21.6
& \\[0.08cm] 
PW Vul    & J & S  & 2.3$^{+0.9}_{-0.5}$ & 0.55(10)
& 2       & 7.0 .. 13.8  & 8     & 0.6 .. 13.8
& \\[0.08cm] 
QU Vul    & P & F  & 1.6$^{+1.7}_{-0.7}$ & 0.55(05) 
& 5       & 2.6 .. 13.5  & 23    & 0.4 .. 13.5
& \\[0.08cm]
CRTS J05  & -- & -- & (1.0$^{+1.2}_{-0.5}$) & (0.32)
& 1       & 483          & 1     & 483
& AN \\[0.08cm]
IGR J17   & -- & -- & 1.01$^{+0.05}_{-0.05}$ & 0.36(12)
& 1       & 600          & --    & --
& AN \\[0.08cm]
IPHASX J21 & -- & -- & 0.73$^{+0.02}_{-0.02}$ & 0.28(10)
& 1       & 147$^{+20}_{-20}$ & 1       & 147$^{+20}_{-20}$
& AN \\[0.08cm]
\hline
\end{tabular}
\end{table*}

In the following, we include short comments on selected systems included in Table \ref{allD01_tab}, particularly concerning the identification of an object with a source in the Gaia Data Release 2 
catalogue. We use the following abbreviations: DR2 for the Gaia Data Release 2 catalogue 
\citep{gaia18-1}, DSS for the Digitized Sky Survey, PS1 for the Panoramic Survey Telescope and 
Rapid Response System \citep[Pan-STARRS,][]{chambersetal16-1,flewellingetal16-1} data
and D05 for the \citet{downesetal05-1} catalogue of cataclysmic variables. Also, the term
`our own observations' refers to the data that will be presented in detail in the forthcoming
second part of this study.

\medskip\noindent
{\em DO Aql:} For the D05 position, DR2 lists two stars with a separation of 0.9 arcsec. The stars are
not clearly resolved in PS1, but our own observations confirm that the western, brighter and
bluer, component is the nova. The coordinates in Table \ref{allD01_tab} correspond to the DR2 source.

\medskip\noindent
{\em V1370 Aql:} The coordinates were determined from our own observations.

\medskip\noindent
{\em V1425 Aql:} The position given in Table \ref{allD01_tab} has been derived from our own 
observations. It places the nova about 11 arcsec off the D05 coordinates. There is no DR2 source
associated with this location.

\medskip\noindent
{\em V812 Cen:} The DR2 source appears to be close to the object marked in D05, but only
DSS images are available at this position, which do not allow for an unambiguous identification.

\medskip\noindent
{\em V868 Cen:} The D05 finding chart marks the nova as the south-western part of a close
visual binary. While the corresponding DSS image does not resolve the binary, the fact that
both components are included in DR2, with a separation of 2.3 arcsec, allows to unambiguously
identify the nova. The position given in Table \ref{allD01_tab}
is taken from the DR2 catalogue.

\medskip\noindent
{\em V888 Cen:}  The D05 coordinates refer to the centroid of a visual binary 
with a separation of 2.3 arcsec, which explains the comparatively large 
difference to the DR2 position. The nova is the southern component, actually being the one of the two 
DR2 sources that is further off the D05 coordinates. The position given in Table \ref{allD01_tab}
is taken from the DR2 catalogue.

\medskip\noindent
{\em IV Cep:} DR2 lists two objects with a separation of 2.4 arcsec. The visual binary is
clearly resolved in the PS1 image, and Table \ref{allD01_tab} lists the bluer of the two.

\medskip\noindent
{\em V450 Cyg:} The finding chart in D05 marks the wrong object. Comparison with 
\citet{slavinetal95-1} and the PS1 images unambiguously identifies the nova.

\medskip\noindent
{\em V476 Cyg:} DR2 lists two objects with a separation of 1.5 arcsec. From the PS1 image, the
south-western, brighter one, is the bluer of the two, and thus more likely to be the nova.
The coordinates in Table \ref{allD01_tab} refer to that DR2 source.

\medskip\noindent
{\em V838 Her:} D05 mark the nova as the fainter, north-western part of a close pair. The PS1 image 
reveals that the marked object is actually also a visual binary aligned at about the same angle
as the original pair. From this line of three stars, only the faintest, most north-western object, 
is not included in DR2. However, because it presents a considerably redder colour than the
central object, the latter DR2 source is more likely to be the nova.

\medskip\noindent
{\em GQ Mus:} The DSS image at the position of the nova is inconclusive, but the DR2 sources
close to that position form a triangle that with respect to both its shape and the brightnesses
of the three objects can be unambiguously identified on the D05 finding chart, with the
southern tip of the triangle being the nova. 

\medskip\noindent
{\em BT Mon:} Coordinates determined from our own observations.

\medskip\noindent
{\em V972 Oph:} DR2 lists two sources with a separation of 1.1 arcsec. The northern object of
the two is not visible in the PS1 image. Our own observations confirm that the nova is the
southern, brighter object.

\medskip\noindent
{\em V2104 Oph:} Coordinates taken from \citet{tappertetal14-1}.

\medskip\noindent
{\em V2214 Oph:} The D05 coordinates point to a string of stars that is insufficiently resolved
on their finding chart. Comparison with PS1 shows that the DR2 source corresponds to the bluest
and brightest object of those. Our own observations confirm that this is indeed the nova.

\medskip\noindent
{\em V2264 Oph:} There is no proper finding chart available on which the nova could be identified.

\medskip\noindent
{\em V4157 Sgr:} There is no proper finding chart available on which the nova could be identified.

\medskip\noindent
{\em V4160 Sgr:} There is no proper finding chart available on which the nova could be identified.

\medskip\noindent
{\em V4169 Sgr:} While D05 do not mark any specific object on their finding chart,
our own observations confirm the DR2 source with the nova.

\medskip\noindent
{\em V4171 Sgr:} There is no proper finding chart available on which the nova could be identified.

\medskip\noindent
{\em V4361 Sgr:} Comparison with PS1 shows that the object marked in the D05 finding chart is
not included in DR2.

\medskip\noindent
{\em V4444 Sgr:} The resolution of the D05 finding chart is insufficient for identifying the
nova in this crowded region. Our own observations show that none of the DR2 sources in the vicinity
of the coordinates correspond to the nova.

\medskip\noindent
{\em V4633 Sgr:} Comparison with PS1 shows that the nova is not in DR2.

\medskip\noindent
{\em V4642 Sgr:} Comparison with PS1 shows that the nova is not in DR2.

\medskip\noindent
{\em V960 Sco:} The finding chart in D05 does not have sufficient resolution to unambiguously
identify the nova. Additionally, their coordinates do not correspond to the location marked
in the chart.

\medskip\noindent
{\em V977 Sco:} The finding chart in D05 does not have sufficient resolution to unambiguously
identify the nova.

\medskip\noindent
{\em V1141 Sco:} A comparison with the D05 finding chart and the PS1 image is somewhat ambiguous,
mainly because of unfortunate marker placement on the D05 chart, but it appears unlikely that
the nova is part of DR2.

\medskip\noindent
{\em V1142 Sco:} Only DSS images are available at that position, and their resolution is
insufficient to identify the nova. There is no DR2 source within 2 arcsec of the D05 position.

\medskip\noindent
{\em FV Sct:} DR2 lists two objects with a separation of 1.9 arcsec near the D05 coordinates.
From the PS1 image, the brighter, north-western object is the bluer one. Our own observations
confirm that this is the nova. The position quoted in Table \ref{allD01_tab} refers to that DR2 source.

\medskip\noindent
{\em V373 Sct:} The DR2 source is confirmed as the nova by our own observations.

\medskip\noindent
{\em V443 Sct:} There is no proper finding chart available on which the nova could be identified.

\medskip\noindent
{\em V444 Sct:} There is no proper finding chart available on which the nova could be identified.
Comparison with PS1 shows that none of the DR2 sources within a radius of 5 arcsec is particularly
blue.

\medskip\noindent
{\em FH Ser:} Coordinates were determined by our own observations.

\medskip\noindent
{\em LW Ser:} While the D05 finding chart is somewhat ambiguous, comparison of the PS1 data with
our own observations show that the nova is not in DR2.

\medskip\noindent
{\em LV Vul:} The D05 finding chart has insufficient resolution in order to identify the nova
out of a blob of at least five stars, all of which are DR2 sources. The coordinates listed
in Table \ref{allD01_tab} were derived from H$\alpha$ imaging obtained by us. Comparison with
PS1 shows that the corresponding DR2 source represents the bluest object in this clump.

\end{appendix}

\end{document}